\documentclass[aps,prper,reprint,superscriptaddress,floatfix]{revtex4-1}

\usepackage{graphicx}
\usepackage{natbib}
\usepackage{color}   
\usepackage[above,below]{placeins}	

\begin{document}
	
\title{Network positions in active learning environments in physics}
	
	
\author{Adrienne L. Traxler}
\email[]{adrienne.traxler@wright.edu}
\author{Tyme Suda}
\affiliation{
	Wright State University, Department of Physics,
	Dayton OH, 45435
}
	
\author{Eric Brewe}
\affiliation{
	Drexel University, Department of Physics,  School of Education
	Philadelphia PA, 19104
}
	
\author{Kelley Commeford}
\affiliation{
	Drexel University, Department of Physics,
	Philadelphia PA, 19104
}
	
\date{\today}
	
\begin{abstract}
This study uses positional analysis to describe the student interaction networks in four research-based introductory physics curricula. Positional analysis is a technique for simplifying the structure of a network into blocks of actors whose connections are more similar to each other than to the rest of the network. This method describes social structure in a way that is comparable between networks of different sizes and densities and can show large-scale patterns such as hierarchy or brokering among actors. We detail the method and apply it to class sections using Peer Instruction, SCALE-UP, ISLE, and context-rich problems. At the level of detail shown in the blockmodels, most of the curricula are more alike than different, showing a late-term tendency to form coherent subgroups that communicate actively among themselves but have few inter-position links. This pattern may be a network signature of active learning classes, but wider data collection is needed to investigate. 
\end{abstract}
	
\maketitle
	
\section{Introduction}
	
Social network analysis (SNA) is a powerful toolkit for understanding the social structure of classrooms. It quantitatively describes the student/student interactions that are foundational to active learning. Network surveys are usable at a class-wide scale even with large enrollments, so they complement qualitative studies that can give deep detail on one or a few student groups. 
Most quantitative methods theoretically position individual traits (concept inventory scores, pass/fail outcomes, etc.)\ as the primary object of interest. However, in the intentionally collaborative and interdependent environment of active learning classrooms, this abstraction misses key ingredients of the learning experience. Network methods place equal theoretical emphasis on the actors and the patterns of connection among them. This dual focus makes networks an excellent lens for studying active learning. 
	
This paper describes and applies a network technique, positional analysis, which is established in sociology but new to physics education research. Using positional analysis, we compare the social positions available in the classroom networks of four research-based introductory physics curricula. Various social network analyses have been done in physics education research (PER), but there is a scarcity of results that compare different institutions and class types using the same surveys and collection methods. Using the multi-site data from the Characterizing Active Learning Environments in Physics (CALEP) project and developing a free implementation of a positional analysis algorithm, we are able to show broad similarities and a few striking differences in the social structure of these active learning environments in physics.

\subsection{Social network analysis in physics education}

	Physics education research and other discipline-based education research (DBER) fields have identified active learning as critical to effective instruction \citep{freeman_active_2014,mastascusa_effective_2011}. Though definitions of active learning are not all precise and aligned, they center interaction with others, particularly other students \citep{meltzer_resource_2012}. Attending to the structure of these interactions in classrooms and other learning environments has led to the use of social network analysis in PER \citep{brewe_roles_2018} and DBER \citep{grunspan_understanding_2014}. SNA has been used in PER in a variety of ways: to identify patterns of participation in an informal learning environment \citep{brewe_investigating_2012}, to investigate discussions \citep{bodin_mapping_2012,bruun_network_2018}, to identify productive online forum discussions \citep{traxler_networks_2018}, to predict grades in current or future classes \citep{bruun_talking_2013,vargas_correlation_2018}, and to predict persistence in physics \citep{zwolak_educational_2018} and in degree programs \citep{forsman_new_2014}.
	
Identifying and describing communities within broader networks is an active research area in SNA \citep{dey_what_2019}. This has not been much used in PER to date, though some descriptive work has looked at the number of detected student communities as a function of time in the semester \citep{bruun_time_2014,traxler_community_2015}. 
Sociologists and other network scientists have developed many community detection algorithms, so the foundation is laid for physics education researchers to analyze classrooms with these tools.

\subsection{Positional analysis}
	
Networks use relational data to quantify the connections between a set of entities. In the case of social networks, these entities (also known as {\em nodes} or {\em actors}) are people, and the structure of relations ({\em edges} or {\em links}) describes the structure of the social group. The underlying idea of positional analysis is to extract this structure, dividing a network into a discrete set of social blocks. Each block is a {\em position}, ``a collection of individuals who are similarly embedded in networks of relations'' \citep[p.\ 348]{wasserman_social_1994}.
	
Positional analysis can be classified as a community detection method \citep[][discussion on block modeling]{newman_structure_2003, fortunato_community_2010}, because it looks for ways to group or block nodes so that they are more like their community members than they are the network as a whole. In many kinds of community detection, link density is the guiding principle, with algorithms finding node groups that are more connected to each other than to the surrounding network \citep{newman_modularity_2006,frank_identifying_1995}. In other cases, ``random walk'' algorithms look for minimal-information ways to describe the nested structure of nodes, communities, and networks \citep{rosvall_maps_2008}. Positional analysis takes a different approach, looking for similarity of connections and grouping together nodes who have the greatest equivalence  \citep{white_social_1976,breiger_algorithm_1975,wasserman_social_1994}. Positional analysis algorithms need some definition of equivalence, a measure of similarity for describing how closely nodes approach equivalence, and a process for maximizing that measure. 
	
\begin{figure}
	\includegraphics[width=0.9\linewidth]{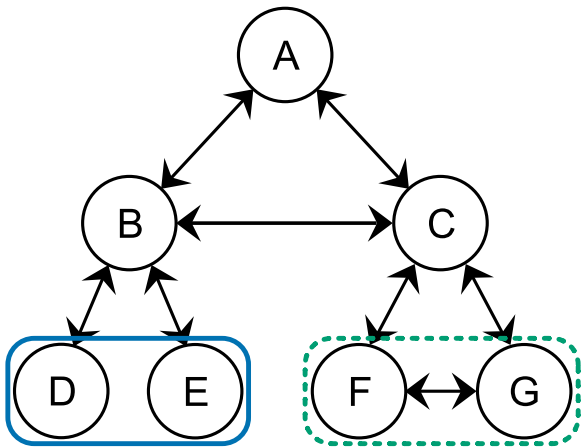}
	\caption{A sample work network, where ties mean ``meets weekly with.'' Positional analysis can identify layers of this example hierarchy, but structural equivalence will produce a more restrictive grouping than regular equivalence. By exact structural equivalence, only D/E (solid blue line) and F/G (dashed green line) are grouped together, with all other actors occupying single-member positions. \label{fig:position-example}}
\end{figure}

Figure \ref{fig:position-example} shows a small example network, which might represent three levels of workers and managers in an organization. Positional analysis can pick out these structural levels as positions, even if not all the people in them are connected to each other (such as nodes D, E, F, and G). The two most commonly used definitions of equivalence are structural and regular. Structural equivalence looks for the {\em same ties to same others} \citep{wasserman_social_1994}. In Fig.\ \ref{fig:position-example}, nodes D and E would be perfectly structurally equivalent (both connected only to node B), as would nodes F and G (both connected to C and to each other). On the other hand, regular equivalence looks for {\em similar ties to similar others}. In Fig.\ \ref{fig:position-example}, nodes B and C might be regularly equivalent, and also nodes D through G (each of which connects to a node in the B/C group). 
	
Regular equivalence is the more conceptually appealing definition in some ways---it can identify the three ``levels'' in Fig.\ \ref{fig:position-example}, while perfect structural equivalence will not group any nodes beyond D/E and F/G. However, regular equivalence is an ill-posed problem, with most networks having many possible partitions that all equally meet the definition of regular equivalence \citep{ferligoj_positions_2011}. Structural equivalence also is less sensitive to missing data \citep{znidarsic_non-response_2012}, which is a common concern in survey-based networks. For these reasons, and because it is mathematically more straightforward to define, we will use structural equivalence to look for network positions. Exact structural equivalence is rare to find in real networks, so methods to calculate it look for the closest possible match \citep{wasserman_social_1994}.

\subsection{Research questions}
	
We will address the following research questions: 
	
\begin{enumerate}
	\item What network positions emerge from the four different curriculum types?
	\item What differences exist between early- and late-term network positions?
	\item What major similarities or differences exist in network positions across learning environments?
\end{enumerate}
	
Section \ref{sec:methods} describes the four classroom settings, the data collection, and the details of the positional analysis algorithm. Section \ref{sec:results} summarizes network characteristics and positional analysis results for the early- and late-semester classroom surveys. Section \ref{sec:discuss} describes common patterns and characterizes the social positions identified by the analysis. Section \ref{sec:conclus} revisits the research questions to give concluding notes, and Sec.\ \ref{sec:future} outlines limitations and future work.

\section{Methods}\label{sec:methods}
	
\subsection{Data and class context}
	
Project sites were chosen for having institutionalized, high-fidelity implementations of research-based curricula in introductory physics. A researcher visited each site to take classroom observation data, discussed in another paper \citep{commeford_characterizing_2020}. In the first and tenth weeks of the term, a survey was distributed to students through Qualtrics which asked them to respond to the following prompt: ``Please choose from the list of people that are enrolled in your physics class the names of any other student with whom you had a meaningful interaction in class during the past week, even if you were not the main person speaking.'' Students were then given a class roster of names to choose from. This form of the survey prompt has been used and validated in previous classroom network studies \citep{zwolak_educational_2018}. Providing a roster and a specific time frame in network surveys boosts recall and reliability of the question (that is, how consistently different students interpret what they are asked) \citep{marsden_survey_2011}. Survey invitations were emailed by the researcher, and reminder emails were sent after 1--3 days.

\subsubsection{Peer Instruction}
	
Peer Instruction \citep{mazur_peer_1997} was developed to allow for active learning in large lecture halls, often with immobile seats. The instructor poses questions with carefully chosen distractors to elicit misconceptions and spark debate. Students individually ``vote'' their answers (using electronic clickers, paper cards, or other means), and the instructor evaluates the answer distribution. In many cases it is useful to have students discuss their answer with one or two neighbors, then the class votes again. Peer Instruction has a long record of evidence for student learning gains \citep{crouch_peer_2007} and is one of the most widely adopted research-based teaching strategies for new physics faculty \citep{henderson_promoting_2008}. 
	
Peer Instruction data were taken from a large, primarily residential private university in the northeastern United States, with a Carnegie classification of Very High Research Activity. The 2017--2018 student body was 24,190 (64\% undergraduate). The racial and ethnic demographics of the students at the university level in Fall 2018 were 8\% African American, 15\% Asian, 6\% Hispanic, 53\% White, 3\% more than one race, $<$1\% Native American and Pacific Islander.

\subsubsection{SCALE-UP}
	
SCALE-UP, short for  Student-Centered Activities for Large Enrollment Undergraduate Programs, is a studio format classroom type where lecture and lab time are combined and computers are on hand to help with activities \citep{beichner_student-centered_2007}. Students work in small groups, typically three groups of three students to a table, on a variety of ``tangible'' or ``ponderable'' activities, labs, or other problem types. Class is usually divided into short segments of 5--15 minutes interspersed with discussions, and lecture occurs among these segments to synthesize and organize the activities. 
	
SCALE-UP data were taken at a large, primarily residential public university in the midwestern United States, with a Carnegie classification of High Research Activity. The 2016--2017 student body was 14,432 (86\% undergraduate). The racial and ethnographic demographics of the students at the university level in Fall 2018 were 3\% Black or African American, 2\% Asian, 2\% Hispanic, 82\% White, 3\% more than one race, 1\% American Indian/Alaskan, $<$1\% Hawaiian/Pacific Islander.

\subsubsection{ISLE}

ISLE, short for Investigative Science Learning Environment, is a curriculum that can be implemented in laboratory sections or (ideally) across the lecture, lab, and recitation components of a class. The units guide students through cycles of observing phenomena, finding patterns, and developing theories to test predictions \citep{etkina_investigative_2007}. There is an emphasis on coordinating multiple representations, and the curriculum uses a cognitive apprenticeship model to help students learn about the nature of science as they develop their physics ideas \citep{etkina_investigative_2007}. The section shown in this paper was a lab-only implementation.

ISLE data were taken from a large, primarily residential public university in the northeastern United States, with a Carnegie classification of Very High Research Activity. The 2018--2019 student body was 42,828 (78\% undergraduate). The racial and ethnographic demographics of the students at the university level in Fall 2017 were 8\% African American, 22\% Asian, 12\% Hispanic, 40\% White, 3\% more than one race, 13\% international, and 2\% other (including Native American and Pacific Islander).

\subsubsection{Context-rich problems}
	
Context-rich problems, also called cooperative problem-solving, is a method often used in recitation sections attached to a lecture course \citep{heller_cooperative_2010,heller_teaching_1992a}. Students work in small groups on problems that are framed less straightforwardly than typical textbook problems (``context rich''), where deciding what quantities must be solved for is often a necessary step. The solution process is explicitly structured to follow expert problem-solving habits \citep{heller_cooperative_2010,heller_teaching_1992a}. Groups have been shown to outperform their highest-scoring individual members, as well as a parallel class taught in a traditional lecture style without this problem-solving framework \citep{heller_teaching_1992a}. 

Context-rich problems data were taken from a very large public associate's college in the western United States. The 2017--2018 student body was 34,642 (100\% undergraduate). The racial and ethnographic demographics of the students in Fall 2017 were 2\% African American, 5\% Asian, 33\% Latino, 50\% White, 6\% more than one race, 3\% unknown, and 1\% Native American and Pacific Islander.

\subsection{Positional analysis with CONCOR}

\subsubsection{The CONCOR algorithm}

The CONCOR (CONvergence of iterated CORrelations) algorithm uses structural equivalence as its basis for sorting the sociomatrix into blocks, with each block corresponding to a social position. The CONCOR loop takes a sociomatrix and performs a Pearson correlation on its columns, iterating until all entries converge to +1 or -1 \citep{breiger_algorithm_1975}. 
These $\pm 1$ values are used to separate the columns into two blocks, and the process can be repeated to further divide blocks. 

An implementation, the concoR package \citep{slez_concor_2015}, already existed for R \citep{R_software}. However, this version of CONCOR was unable to reproduce known results for Krackhardt's High-tech Managers network \citep[][page 379]{wasserman_social_1994}. 
Under further scrutiny it was found that the concoR package only considered outgoing and not incoming ties, and did not appropriately exclude self-ties from the calculations. This form of the algorithm works similarly to the original version described in \citet{breiger_algorithm_1975}, but the more modern form uses both rows and columns (incoming and outgoing ties) to correlate links.
We could not find another fully-functioning open source version of CONCOR, so we wrote our own. Our version, available online \citep{suda_CONCOR_2019} and to be submitted for review elsewhere, incorporates information about both tie directions and successfully reproduces Wasserman and Faust's results on the High-tech Managers network data \citep[][ch.\ 9]{wasserman_social_1994}.

The Pearson correlation between nodes $i$ and $j$ is defined as follows for multiple relations \citep{wasserman_social_1994}:

\begin{widetext}
	\begin{equation}\label{eq:pearson}
	r_{ij} = \frac{\sum_{r=1}^{2R}\sum_{k=1}^{N} (x_{ikr}-\overline{x}_{i\bullet}) (x_{jkr}-\overline{x}_{j\bullet})} {\sqrt{\sum_{r=1}^{2R}\sum_{k=1}^{N}(x_{ikr}-\overline{x}_{i\bullet})^2} \sqrt{\sum_{r=1}^{2R}\sum_{k=1}^{N}(x_{jkr}-\overline{x}_{j\bullet})^2}}.
	\end{equation}
\end{widetext}

\noindent Here $R$ denotes the total number of relations (liking, animosity, etc.) between $N$ nodes. Sums in (\ref{eq:pearson}) are for $i \neq k$, $j \neq k$, to avoid counting self-ties. The mean of values in row $i$ is $\overline{x}_{i\bullet}$, and $\overline{x}_{\bullet i}$ is the mean of values in column $i$ of the $N \times N$ sociomatrix. In the original formulation of CONCOR, correlations were to be run on the columns of the sociomatrix, with the possibility mentioned of correlating between rows instead \citep{breiger_algorithm_1975}. The later version of the correlation calculation given by \citet{wasserman_social_1994} uses the information in both rows and columns (both outgoing and incoming ties) by appending the transpose of the matrix to the original sociomatrix and then correlating the columns \citep{burt_positions_1976}. 

An isolated node---one with no incoming or outgoing ties---should be structurally equivalent to other isolates, as they have the same connections (none) to all other nodes. The correlation loop in CONCOR fails for matrix columns of all zeros, so our version identifies and blocks the isolates together before running correlations on the rest. Isolates are reported as a separate block at the end of the CONCOR-identified blocks. After this change, our CONCOR implementation worked for an arbitrary number of splits (assuming there are structurally inequivalent nodes to separate) and allows for an arbitrary number of different relations to be included. A single measured network (for example, meaningful interactions in week 1) is a single relation, but two time points for the same network could be construed as multiple relations and used to show the time development of positions \citep{white_social_1976}. In the results below, we calculated single-relation CONCOR blocks for each network sample (week 1 and week 10 for each class).

\subsubsection{Treatment of missing data}

Missing data is a common problem in network analyses, and can bias results by altering the network structure being studied. There are several common methods for handling this issue: reporting rates but otherwise ignoring missing data, restricting the network to ties between respondents, imputing missing data, and correcting via exponential random graph models \citep{huisman_imputation_2009,smith_structural_2013}. \citet{huisman_imputation_2009} simulated the effects of missing data and several imputation methods on a friendship network of 50 actors, which is in the same size and density range as the classroom networks in our study. He found that for directed networks with more than small amounts (30\%) of missing data, imputing the unobserved ties gave more biased estimates of network statistics than ignoring missingness for most measures. In another study, \citet{znidarsic_non-response_2012} simulated the effects of missing data on blockmodels of small networks. They found that overall, restricting to complete cases (dropping non-respondents and any reported links to them) caused the least distortion in identifying network positions. After reviewing these results, we opted to avoid imputation and report results for two versions of the networks: all observed ties (ignoring missingness in the data), and complete cases only (removing non-respondent nodes and any links to them).

\subsubsection{Response rates}
	
\begin{table*}
	\caption{Survey response rates and proportions of observed and unobserved ties for the sections analyzed in this paper. \label{tab:responses}}
	\begin{ruledtabular}
	\begin{tabular}{llccccccc}
		& & Nodes & Responded & Response & & Partially & Missing & Missing \\
		Site & Time & ($n$) & ($1-m$) & rate & Observed & observed & (NR--R) & (NR--NR)\\ 
		\hline
		Peer Instruction & Early & 116 & 81 & 70\% & 0.49 & 0.21 & 0.21 & 0.09 \\
		Peer Instruction & Late & 116 & 76 & 66\% & 0.43 & 0.23 & 0.23 & 0.12 \\
		SCALE-UP & Early & 71 & 29 & 41\% & 0.16 & 0.25 & 0.25 & 0.35 \\
		SCALE-UP & Late & 71 & 41 & 58\% & 0.33 & 0.25 & 0.25 & 0.18 \\
		ISLE & Early & 27 & 14 & 52\% & 0.26 & 0.26 & 0.26 & 0.22 \\
		ISLE & Late & 27 & 16 & 59\% & 0.34 & 0.25 & 0.25 & 0.16 \\
		Context-rich problems & Early & 48 & 27 & 56\% & 0.31 & 0.25 & 0.25 & 0.19 \\
		Context-rich problems & Late & 48 & 20 & 42\% & 0.17 & 0.25 & 0.25 & 0.36
	\end{tabular}
	\end{ruledtabular}
\end{table*}          

We chose sections for analysis with the highest possible response rates while still representing a range of curricula. If $n$ is the number of possible actors (students in a section) and $m$ is the number who did not respond to the survey, then the actor response rate is $(1-m/n)$ \citep{znidarsic_non-response_2012}. The $(n-m)$ respondents will have $(n-m)(n-m-1)$ measured ties among each other with values of either 0 or 1. These are the fully observed ties, and their proportion to all possible ties in the network is $(n-m)(n-m-1)/n(n-1)$. Partially observed ties connect respondents and non-respondents, and their proportion of the total is $(n-m)m/n(n-1)$. There are two categories of missing ties: those from non-respondents to respondents (fraction $m(n-m)/n(n-1)$) and those between non-respondents (fraction $m(m-1)/n(n-1)$). Table \ref{tab:responses} shows the sections chosen for analysis and their proportions of observed and unobserved ties. The best-case data displayed displayed here is for Peer Instruction (66--70\% of ties partially or completely reported) and the worst case is SCALE-UP (41--58\% of ties partially or completely reported).

\subsubsection{Outputs}

The first iteration of CONCOR on a matrix will divide it into two sets of columns, representing two groups of nodes whose edges are more similar to each other than to the remaining nodes. Running CONCOR again will further divide each of those groups into two subgroups. The algorithm stops after it has reached a specified number of ``splits.'' The appropriate number of splits depends on the application and often must be found experimentally. In the case of our data, one split tended to produce large positions, which were not very informative. Three or more splits often failed, because the algorithm will not converge if it tries to divide a completely-connected group. We report results for two CONCOR splits for all sections. 

The outputs of positional analysis, whether by CONCOR or another algorithm, are:
\begin{itemize}
	\item A {\em blockmodel}, which permutes the rows and columns of the original adjacency matrix to partition the network into positions.
	\item An {\em image matrix}, which shows only the edge density within each block.
	\item A {\em reduced network}, which thresholds values in the image matrix to 0 or 1 and plots the result as a network.
\end{itemize}
Each level of output further condenses the information in the original network, with the reduced network plot showing a graphical summary of the network positions and the connections between them. For calculating the density matrix, each block (whether on- or off-diagonal) has a density equal to its number of 1's divided by the number of possible links in it (number of rows $\times$ number of columns in the block). Self-ties are omitted, so for on-diagonal blocks with $n$ nodes, the number of possible ties is $n(n-1)$. 

\begin{figure*}
	\includegraphics[width=0.9\linewidth]{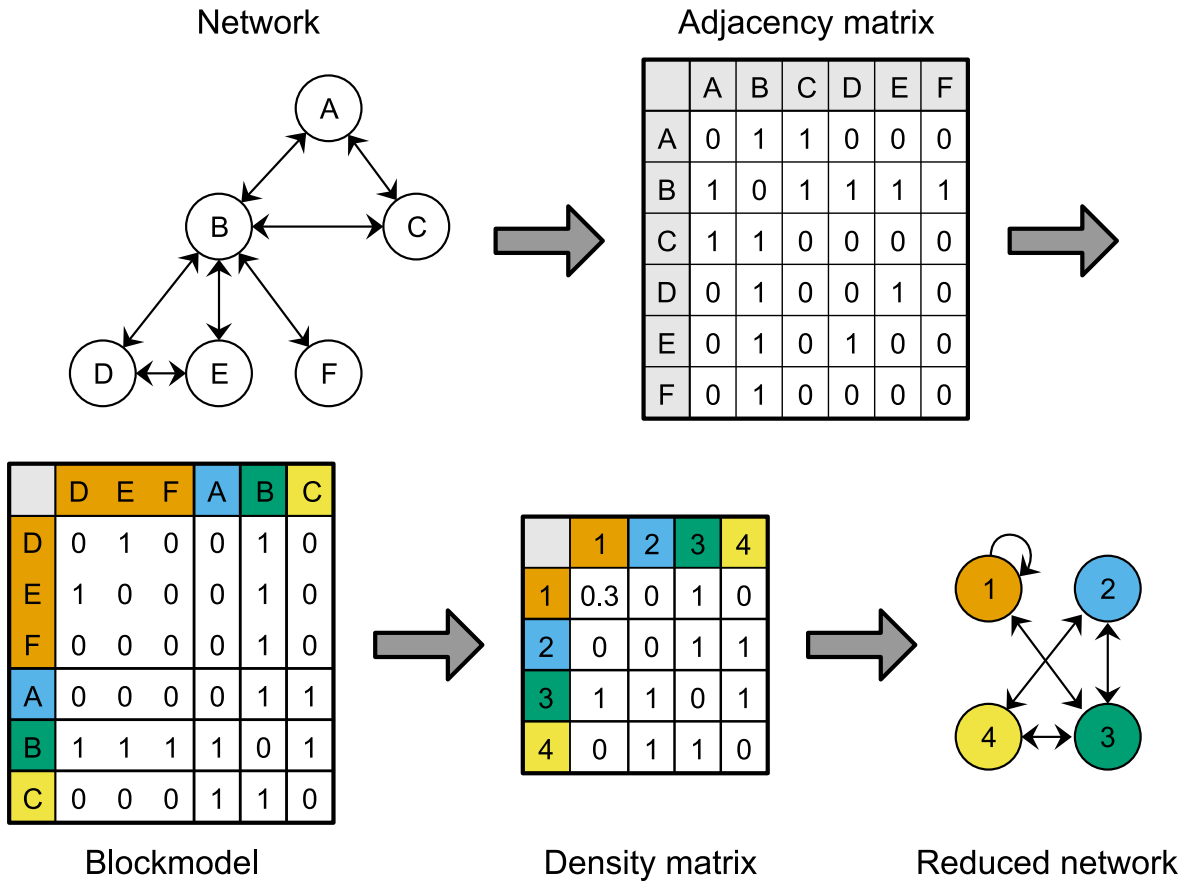} \\
	\caption{Sample of blockmodel outputs. The network (top left) can be written as an adjacency matrix showing which nodes are linked (top right). The bottom row shows the blockmodeling products: a permuted adjacency matrix (left), the matrix showing each block density (center), and a reduced network where each diagonal block is treated as a node and inter- or intra-block links are present or absent according to a density threshold. \label{fig:bm-outputs}}
\end{figure*}

Figure \ref{fig:bm-outputs} shows the stages of the process for a fictional network. We will show the original network, the blockmodel, and the reduced network for each class section and time point.

\section{Results}\label{sec:results}

\subsection{Network descriptions}

Table \ref{tab:nwdesc} shows descriptive statistics for the networks discussed in this paper. This includes number of nodes and edges (including non-respondents who were named by survey takers), reciprocity among respondents, network density, average degree, and clustering coefficient. The Supplemental Material \citep{suppMat} gives the same results for the complete-cases networks, where only survey respondents and links among them are retained. 

Reciprocity is calculated as the probability that for each directed edge, the opposite-pointing edge is also in the graph. Density, average degree, and transitivity are three related ways of describing how well connected a network is. Density is the fraction of present to possible ties, and for a directed network, is 
\begin{equation}
\rho = N_e/N(N-1)
\end{equation}
where $N$ is the number of nodes and $N_e$ is the number of edges. 
Because density scales as $1/N^2$, it is misleading to compare densities between networks of different sizes (larger networks will tend to have lower density). The last two measures in the table give estimates of connectivity that are comparable across networks of different sizes. Average degree describes nodes' number of outgoing and incoming connections, and is calculated by averaging the row and column sums of the adjacency matrix. Transitivity, or clustering coefficient, is the probability that any two nodes who have a neighbor in common will themselves be connected (``the friend of my friend is also my friend'') \citep{newman_structure_2003}. All statistics are calculated in R \citep{R_software} using the igraph package \citep{igraph}. 

\citet{snijders_non-parametric_1999} outline a bootstrap process for estimating the variability in network statistics: resample the network many times, recalculate the statistic using each sampled network, and then calculate the standard deviation of all the sampled statistic values. This estimates the standard error for the measured value of the statistic. Table \ref{tab:nwdesc} includes standard errors for density, average degree, and transitivity using the method of ref.\ \citealp{snijders_non-parametric_1999} with 1000 bootstrap trials. 

\begin{table*}
	\caption{Network descriptive statistics for the classes analyzed: number of nodes ($N$) and edges ($N_e$), fraction of named ties that were reciprocated, network density, average degree, and transitivity. For the last four values, the standard error of the final digit is given in parentheses.\label{tab:nwdesc}}
	\begin{ruledtabular}
	\begin{tabular}{llrrrrrrr}
		Site & Time & $N$ & $N_e$ & Reciprocity & Density & Avg.\ Degree & Transitivity\\ 
		\hline
		Peer Instruction & Early & 94 & 130 & 0.63 & 0.015(3) & 2.8(3) & 0.23(9) \\ 
		Peer Instruction & Late & 97 & 159 & 0.70 & 0.017(3) & 3.3(3) & 0.23(9) \\ 
		SCALE-UP & Early &  56 &  68 & 0.60 & 0.022(6) & 2.4(3) & 0.3(1)  \\ 
		SCALE-UP & Late &  66 & 176 & 0.65 & 0.041(7) & 5.3(5) & 0.52(9) \\ 
		ISLE & Early &  20 &  19 & 0.50 & 0.05(2) & 1.9(4) & 0.3(2)  \\ 
		ISLE & Late &  24 &  47 & 0.93 & 0.09(2) & 3.9(5) & 0.6(2)  \\ 
		Context-rich problems & Early &  40 &  48 & 0.42 & 0.031(8) & 2.4(3) & 0.2(1) \\ 
		Context-rich problems & Late &  41 &  96 & 0.62 & 0.06(2) & 4.7(6) & 0.24(9) \\ 
	\end{tabular}
	\end{ruledtabular}
\end{table*}

The Peer Instruction network shows only a marginal increase in density over the semester, but in the other three classes, the week 10 density roughly doubles from its week 1 value. A similar pattern exists for average degree. For transitivity, the Peer Instruction and context-rich problems values stay relatively steady, while the other two class types increase. 

In the Peer Instruction and SCALE-UP sections, the majority of links are reciprocated in both week 1 and 10. In ISLE, the week 1 reciprocity rate is lower at 50\%, but in week 10 is the highest, at 90\%. Context-rich problems show a smaller increase, with the lowest starting reciprocity but a week 10 value comparable to Peer Instruction and SCALE-UP. There is no clear pattern connecting class size with reciprocity, average degree, or transitivity. 
	
	
\subsection{Peer Instruction}  
	
\begin{figure*}
	{\centering
		\includegraphics[width=0.9\linewidth]{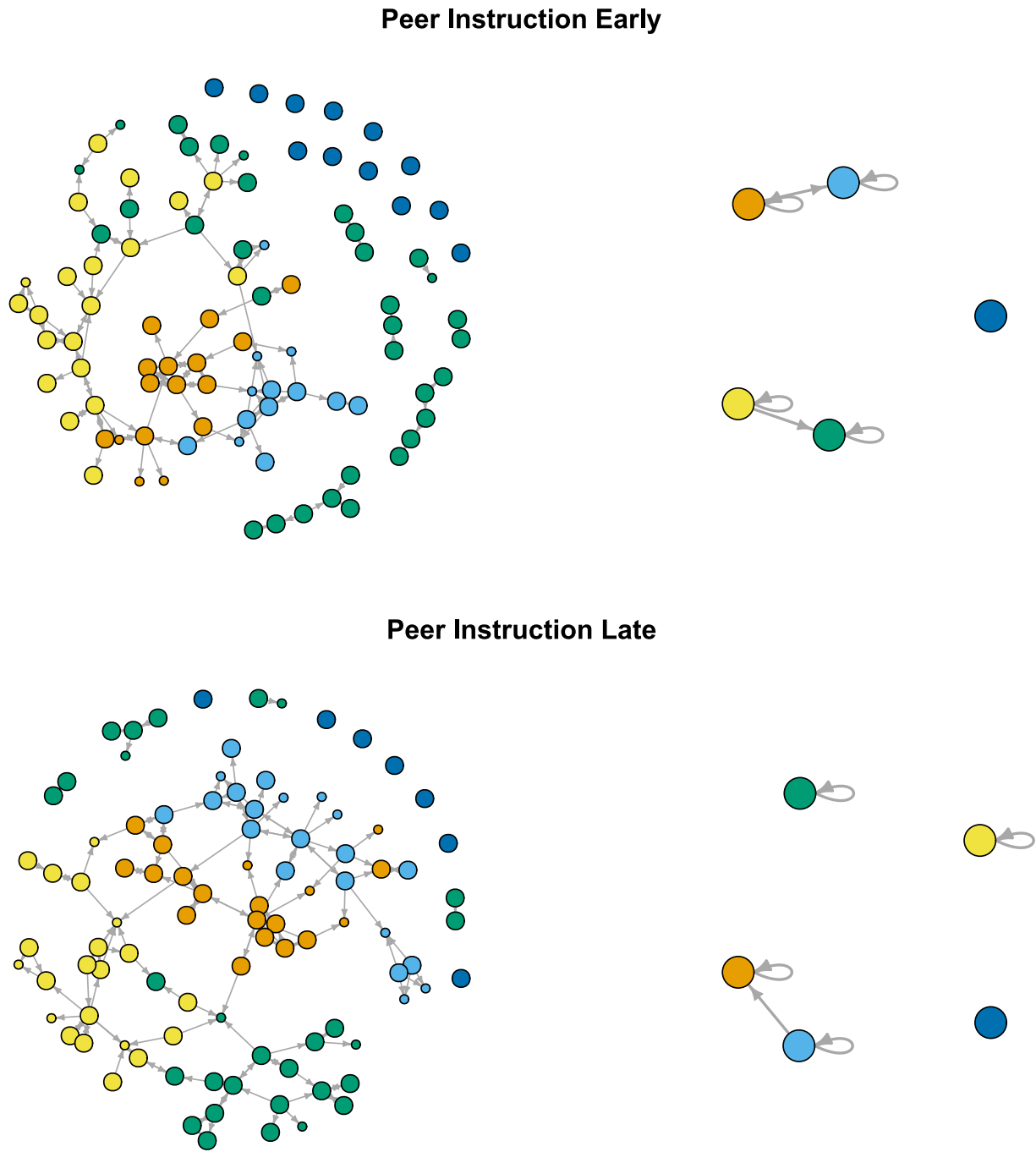} }
	\caption{Week 1 (top) and week 10 (bottom) network and reduced network diagrams for Peer Instruction section. The left column shows the sociogram of students who took the survey, plus those who did not take the survey but were named by respondents (smaller circles). Nodes are colored by CONCOR partition. The right column shows the corresponding reduced networks, where each circle represents a position from the blockmodel and stands for all of the nodes of the same color in the sociogram. Links in the reduced network plots show connections within and between positions and come from applying a density threshold to the blockmodel. \label{fig:PI-networks}}
\end{figure*}

The left column of Fig.\ \ref{fig:PI-networks} shows the sociograms (network diagrams) for week 1 and week 10 of a Peer Instruction section. The larger circles are survey respondents, while smaller circles are named non-respondents. The links are directed, so only a double-headed arrow between two nodes is reciprocal. Nodes are colored by their position (the partition they were assigned by CONCOR, which can change from week 1 to week 10). In week 1, the giant component is comprised of the first two positions (orange and light blue nodes), the fourth position (yellow), and some nodes in the third position (green). Largely, however, green nodes are in smaller clumps not linked to the giant component. From week 1 to week 10, more of the nodes become linked to the giant component, but there are still several smaller groups and seven isolates (compared to 12 in week 1). 

The right column of Fig.\ \ref{fig:PI-networks} shows the corresponding reduced networks. Four of the positions have substantial communication among themselves both early and late-semester (represented by the self-connected ``loop'' edges in the right column of Fig.\ \ref{fig:PI-networks}). Early in the semester, there appears to be more cross-talk between positions, with one additional link between positions four and three, and another between positions one and two.

\begin{figure*}
	\includegraphics[width=0.95\linewidth]{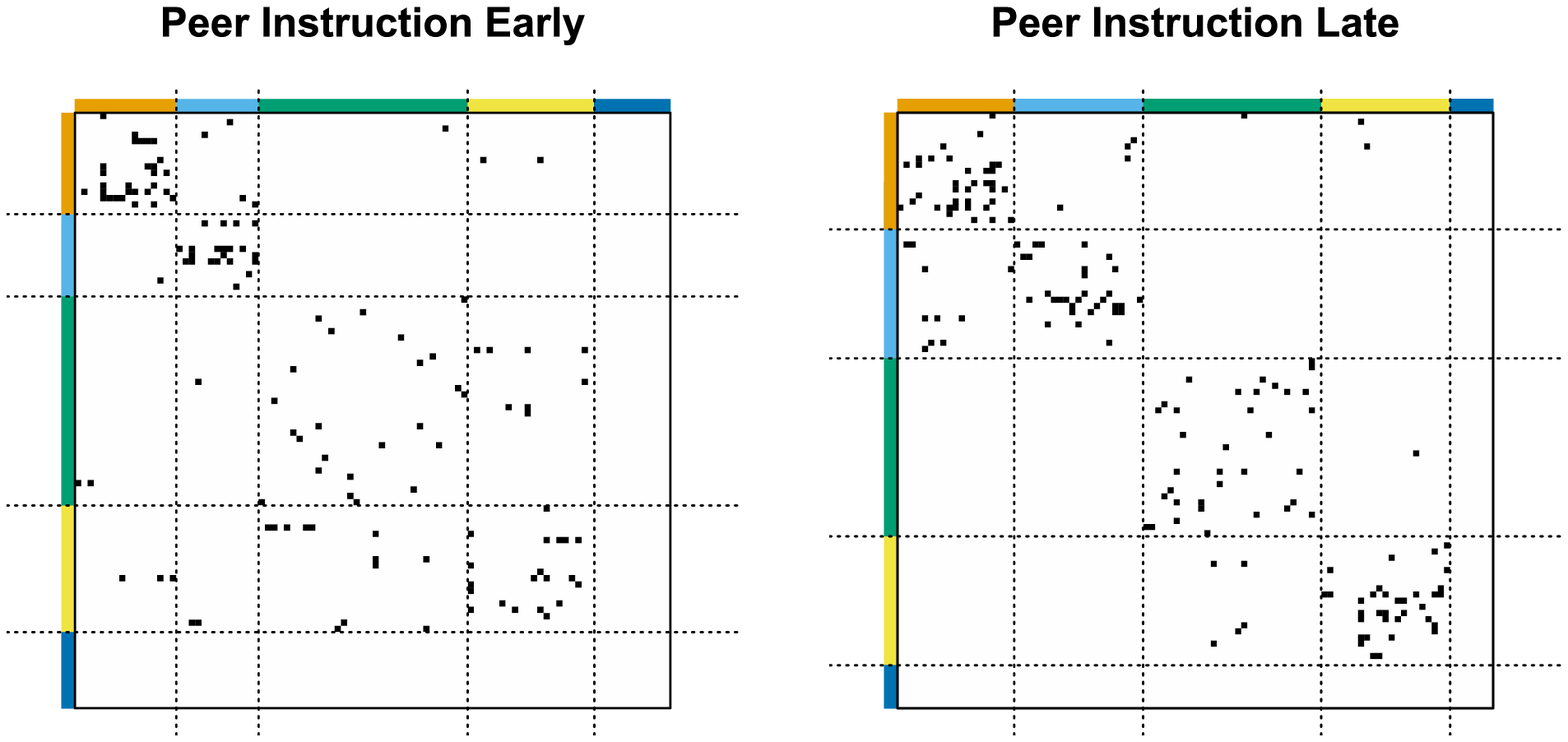} 
	\caption{Week 1 (left) and week 10 (right) blockmodel for Peer Instruction section. Each link from the survey is a black square and would be a `1' rather than a `0' if the adjacency matrix was plotted as numbers. The dotted lines mark the CONCOR group partitions. Colors along the sides mark which blocks of nodes belong to each partition on Fig.\ \ref{fig:PI-networks}. \label{fig:PI-BM}}
\end{figure*}

Figure \ref{fig:PI-BM} shows the blockmodel for the same section, which is the adjacency matrix permuted to group together CONCOR positions in blocks. The blockmodel plots contextualize the change in reduced networks from early- to late-semester. In week 1, the first two positions (the top two on-diagonal blocks in Fig.\ \ref{fig:PI-BM}) are notably higher-density than others. The second two positions show CONCOR's attempt to segregate a more diffuse collection of links. In week 10, the on-diagonal blocks have a more similar density. 

\subsection{SCALE-UP}
	
\begin{figure*}
	{\centering
		\includegraphics[width=0.9\linewidth]{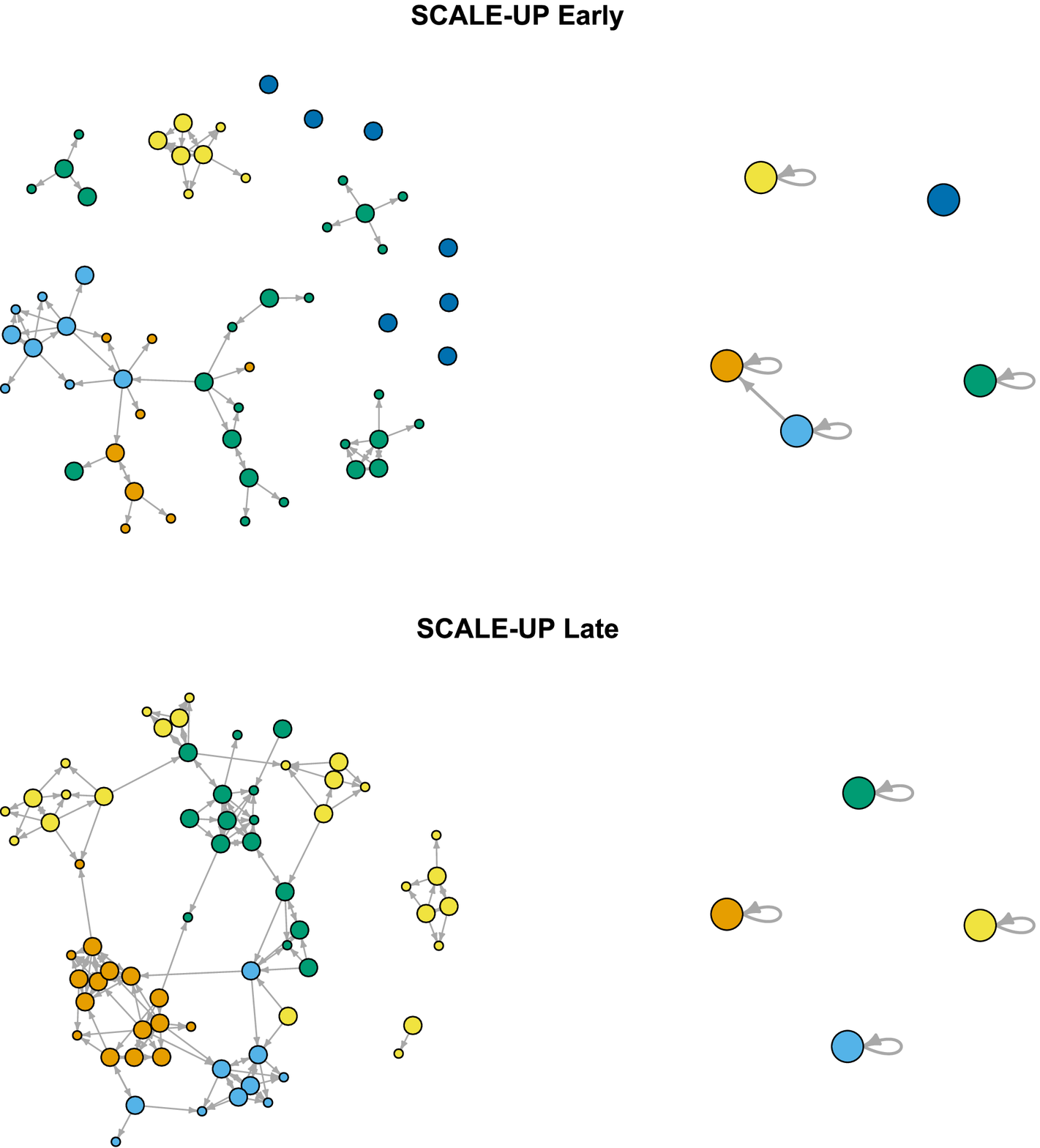} }
	\caption{Sociograms (left column) and reduced network diagrams (right column )for SCALE-UP section. Week 1 is shown in the top row, and Week 10 is the bottom row. Smaller circles on the sociograms are students who did not take the survey but were named by respondents. Nodes are colored by CONCOR partition. \label{fig:SU-networks}}
\end{figure*}

Figure \ref{fig:SU-networks} shows the sociograms and reduced networks for week 1 and week 10 of the SCALE-UP section. There is a significant increase in link density from week 1 to week 10, and all late-term survey respondents reported at least one meaningful interaction (no more isolates exist on the graph). 
In week 1, only 48\% of the nodes are in the giant component, which includes all of positions one and two (orange/light blue) and some of position three (green). Position four (yellow) is a fairly coherent subgroup of seven nodes. 
The week 10 CONCOR results divide most of the giant component, which now has 88\% of the nodes, into three subgroups that are fairly coherent among themselves. The fourth position (diagonal block four in the blockmodel) is a collection of relatively peripheral nodes. 

\begin{figure*}
	\includegraphics[width=0.95\linewidth]{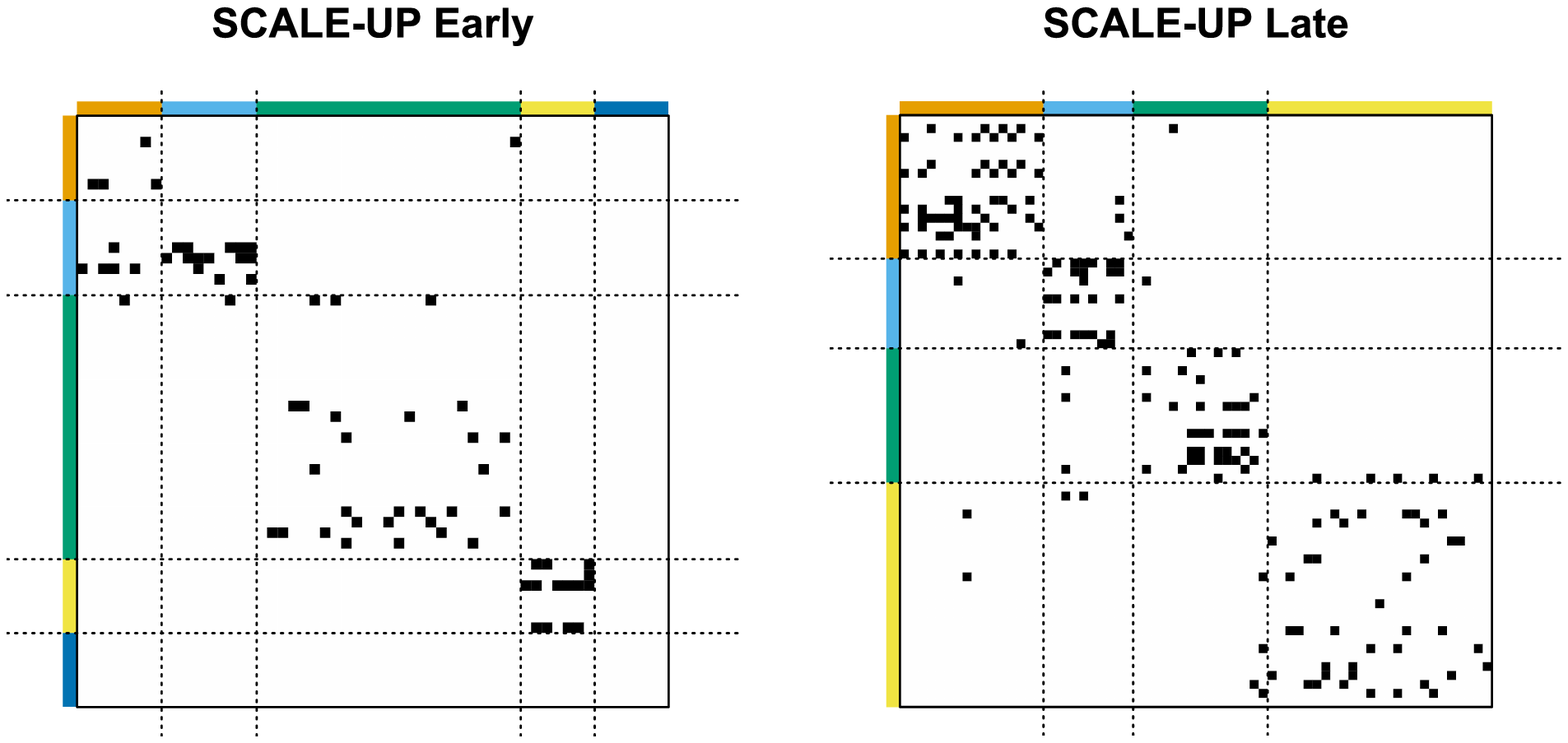} 
	\caption{Week 1 (left) and week 10 (right) blockmodels for SCALE-UP section. Each link from the survey is a black square. The dotted lines mark the CONCOR group partitions. \label{fig:SU-BM}}
\end{figure*}

Figure \ref{fig:SU-BM} shows the blockmodel plots. In week 1 there were three small positions, two of them fairly coherent, and a larger group of nodes with less interconnection. This group (third diagonal block) corresponds to the green nodes on the graph, many of whom were named by survey respondents but did not take the survey themselves (and thus links they may have are unknown). By week 10, the positions are more uniform in size and higher density. On the reduced graphs, both early and late-term, all positions except the isolates (dark blue) talk among themselves. In week 1, there is also an appreciable amount of interaction from position two to one (light blue $\rightarrow$ orange), but this is gone in week 10. 


\subsection{ISLE}

\begin{figure*}
	{\centering
		\includegraphics[width=0.9\linewidth]{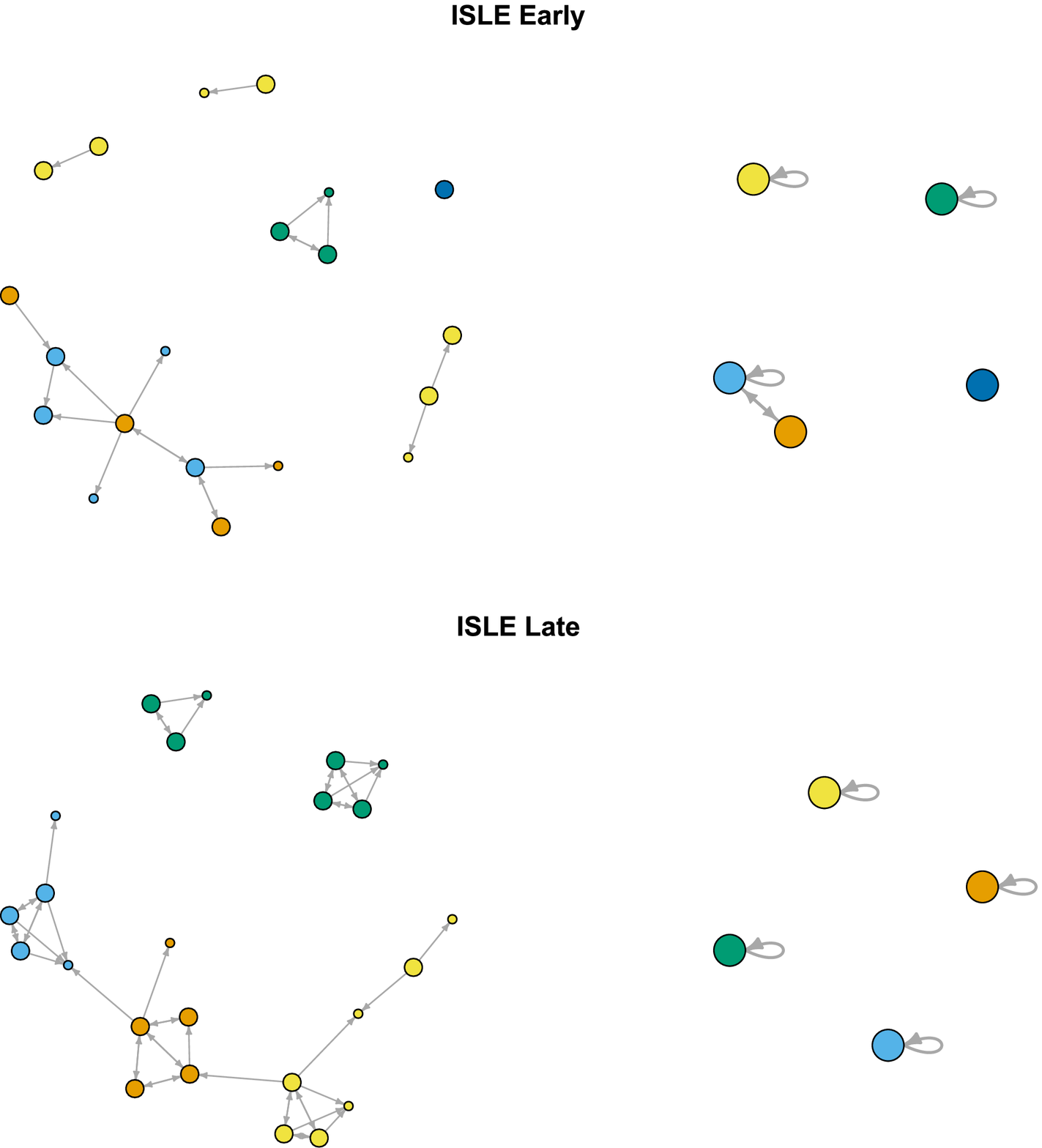} }
	\caption{Week 1 (top) and week 10 (bottom) sociograms (left column) and reduced network diagrams (right column) for ISLE section.\label{fig:IS-networks}}
\end{figure*}

\begin{figure*}
	\includegraphics[width=0.95\linewidth]{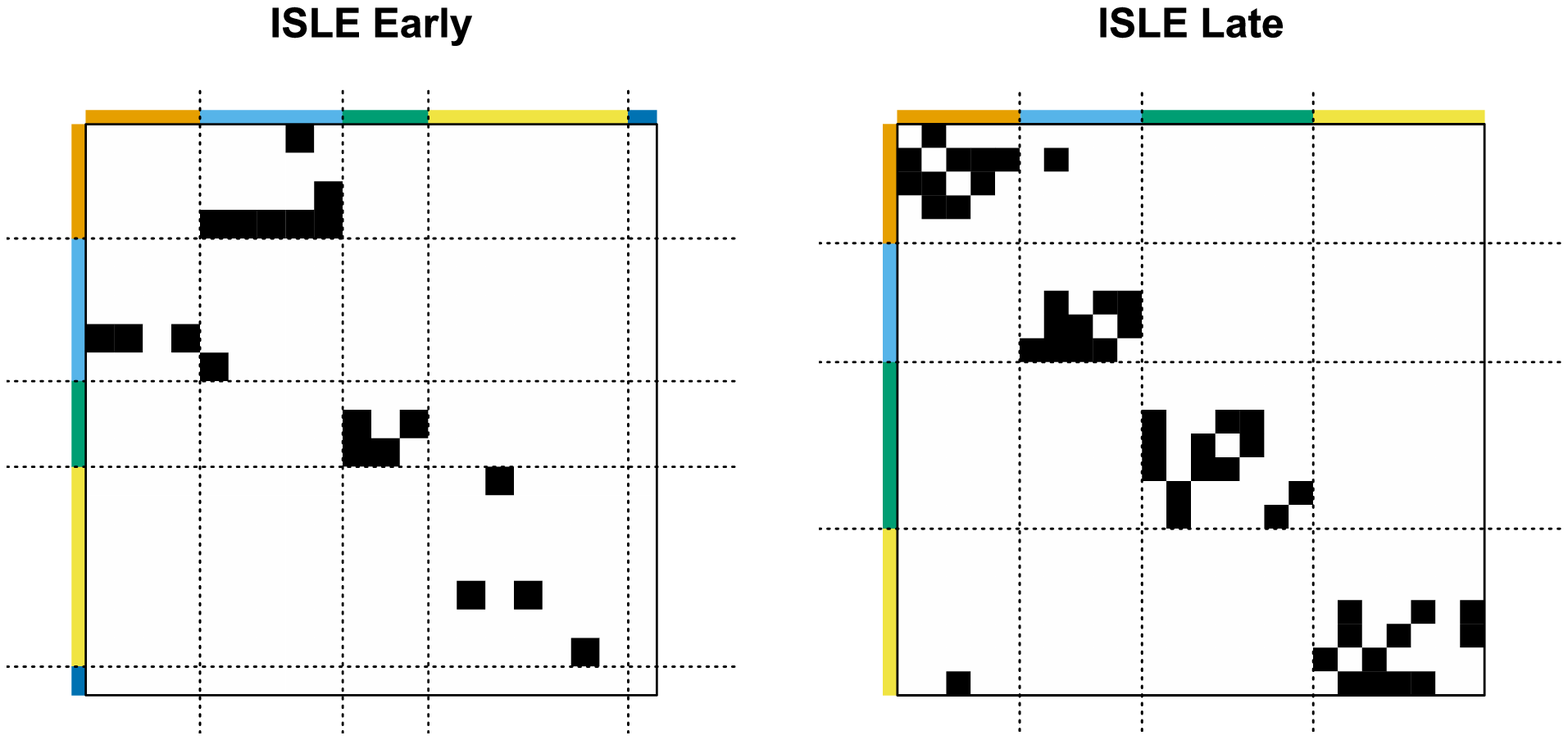} 
	\caption{Week 1 (left) and week 10 (right) blockmodels for ISLE section. Each link from the survey is a black square. The dotted lines mark the CONCOR group partitions. \label{fig:IS-BM}}
\end{figure*}

Figure \ref{fig:IS-networks} shows the sociograms and reduced networks for week 1 and week 10 of the ISLE section. Figure \ref{fig:IS-BM} shows the blockmodels. In week 1, the network is low-density, with fewer than half of the nodes in the giant component. In week 10, that component is larger (71\% of nodes), and is now divided between three of the four CONCOR-identified positions. This relatively clean separation translates into a reduced network with four "island" positions, who talk among themselves but not much between positions. 

Looking at the sociogram and blockmodel for week 1 shows some of the differences between positional analysis and most community-funding algorithms. Light blue nodes connect to orange and vice versa, but are only adjacent to nodes in their own position in one case (the single light blue $\rightarrow$ orange link on the week 1 blockmodel). Community detection routines do not generally ``skip'' in-between nodes to cluster two or more nodes that do not link to each other. 


\subsection{Context-rich problems}

\begin{figure*}
	{\centering
		\includegraphics[width=0.9\linewidth]{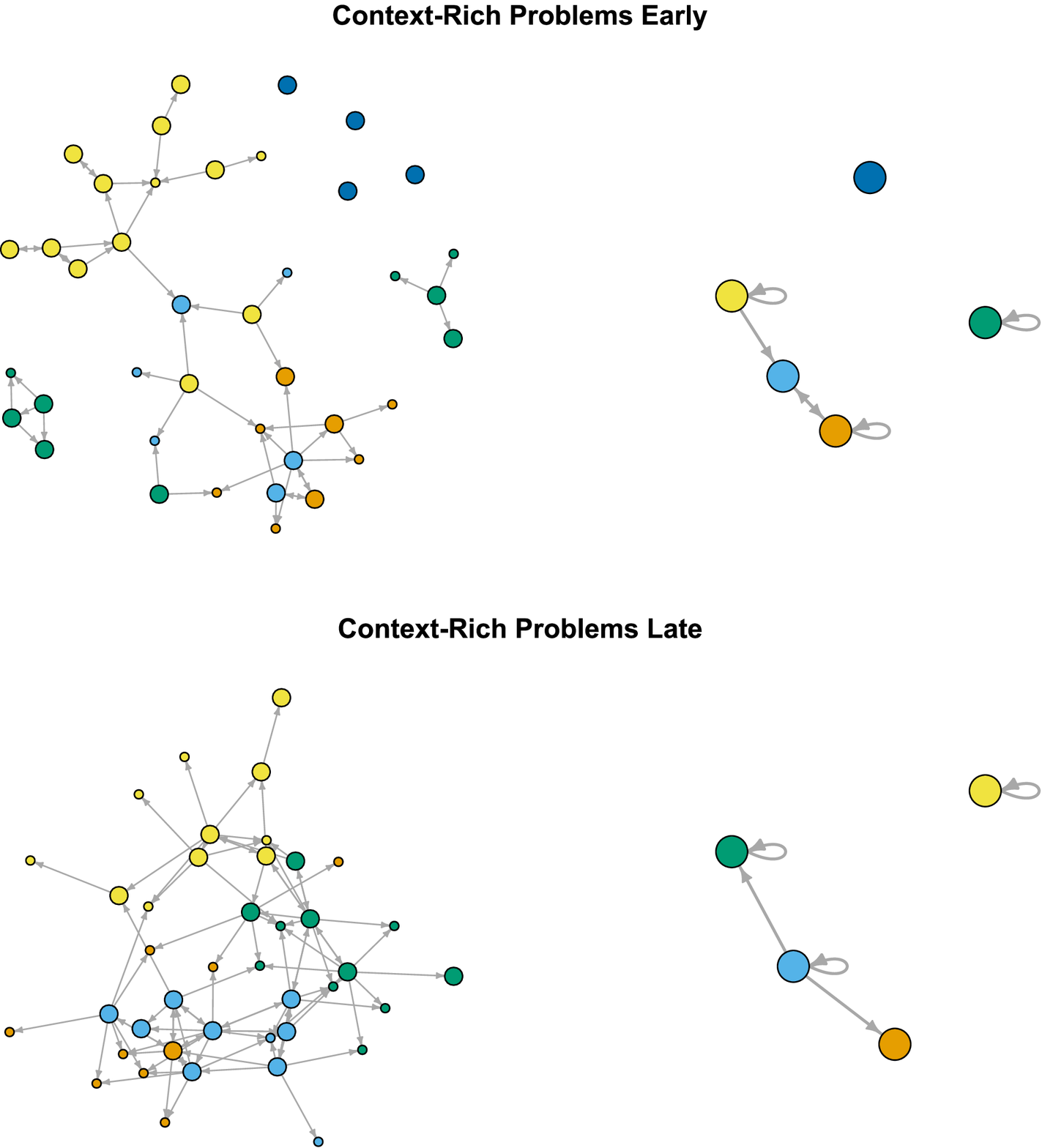} }
	\caption{Week 1 (top) and week 10 (bottom) sociograms (left column) and reduced network diagrams (right column) for context-rich problems section.\label{fig:CR-networks}}
\end{figure*}

\begin{figure*}
	\includegraphics[width=0.95\linewidth]{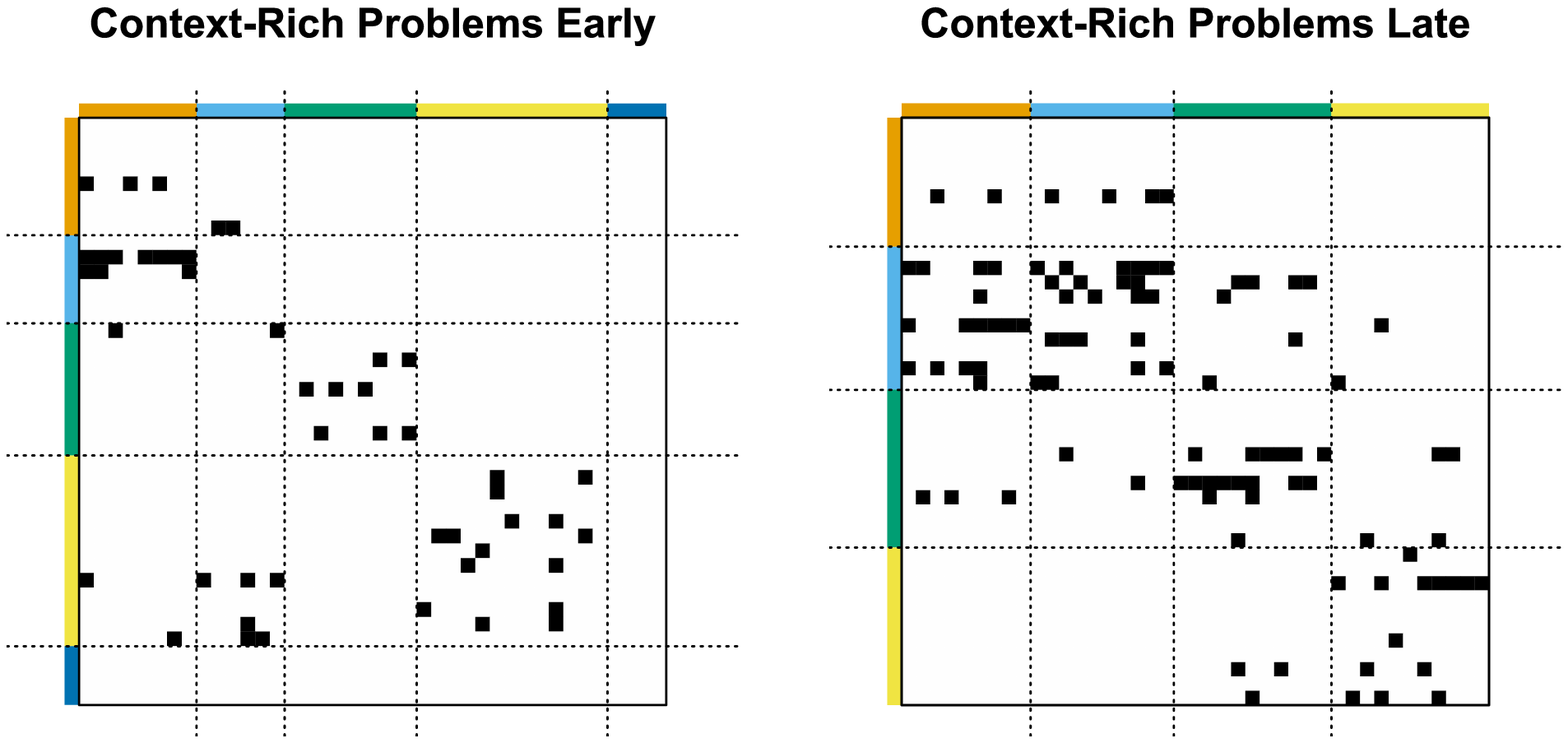} 
	\caption{Week 1 (left) and week 10 (right) blockmodels for context-rich problems section. Each link from the survey is a black square. The dotted lines mark the CONCOR group partitions. \label{fig:CR-BM}}
\end{figure*}

Figure \ref{fig:CR-networks} shows the sociograms and reduced networks for week 1 and week 10 of the context-rich problems section, and Figure \ref{fig:CR-BM} shows the blockmodels. In week 1 (and as in other early-term classes), the network is relatively sparse, with a loosely-connected giant component, two ``island'' groups, and several isolates. The week 10 network has doubled in density, with all respondents in the giant component and a nearly 50\% increase in reciprocity in naming interactions. The reduced network has the highest number of links between positions of any class analyzed, both early and late in the semester. 

The low response rate in week 10 appears to cause some artifacts in the CONCOR blocking. In particular, position one (orange) is dominated by non-respondents, and likely would have more outgoing links if those people had taken the survey. The other positions, particularly three and four, have a lesser version of the same problem. 

\section{Discussion}\label{sec:discuss}

Blockmodels can be interpreted by validating against actor attributes, by describing the overall blockmodel, and by detailing the individual positions \citep{wasserman_social_1994}. Actor information is not available for most of our data, but might include demographic details, grades or other learning outcomes, or schedule information such as students' lab sections \citep{bruun_time_2014}. We can describe the overall blockmodel and individual positions even without additional data.

\subsection{Describing the blockmodel}

A large-scale view of the network structure can come from the image matrix, which summarizes the blockmodel as an adjacency matrix of 1's and 0's between and within blocks. The reduced network plots represent the image matrix visually. Blocks are thresholded to 0 or 1 by comparing their internal density values to a single value $\alpha$ for the network. We use the density of the full network as the threshold value $\alpha$ \citep{wasserman_social_1994}. For example, a three-person block would have six possible ties among members, and a density of 0.5 if three of those ties were actually present. A network density of $\alpha = 0.1$ would mean that the three-person block had an image matrix value of 1 for its self-tie.

Image matrices can be compared to ideal types as a limiting case. Ideal patterns include cohesive subgroups, center-periphery, centralized, and others \citep{wasserman_social_1994}. In the cohesive subgroups pattern, actors in one block primarily talk to each other and not to members of other blocks, corresponding to an image matrix with 1's on the diagonal and 0's elsewhere:
\[
\left[ \begin{array}{cccc}
1 & 0 & 0 & 0 \\
0 & 1 & 0 & 0 \\
0 & 0 & 1 & 0 \\
0 & 0 & 0 & 1 
\end{array} \right]
\]
In a centralized network, all positions link to one other position, whose members also talk among themselves:
\[
\left[ \begin{array}{cccc}
1 & 0 & 0 & 0 \\
1 & 0 & 0 & 0 \\
1 & 0 & 0 & 0 \\
1 & 0 & 0 & 0 
\end{array} \right]
\]
See chapter 10 in Ref.\ \citealp{wasserman_social_1994}, or Ref.\ \citealp{white_social_1976} for more detailed examples. 

Most of the class networks do not perfectly match any of the ideal types. In two cases---late-semester SCALE-UP and ISLE---the cohesive subgroups pattern is exactly present. More commonly, the image matrices largely resemble cohesive subgroups, with one or two links ``extra'' or a self-tie missing. The image matrices for context-rich problems display the least resemblance to any of the ideal types, but also had the lowest response rates. It is possible that a more regular structure was simply not observed.

\subsection{Describing individual positions}\label{subsec:positions}

Each position in the blockmodel can also be considered in light of how it connects to the others and its ratio of communication inside vs.\ outside the block \citep[][chapter 10]{wasserman_social_1994}. If there are $N$ nodes in the network, and $N_k$ in position $k$, there are $N_k(N_k-1)$ ties possible within the position. Collectively, the position has $N_k(N-1)$ ties possible with the whole network (including itself). We can quantify a block's tendency toward self-interactions by comparing to the ratio of internal to total ties. 
For blocks with no internal vs.\ external preference, this ratio is:

\begin{equation}
\frac{N_k(N_k-1)}{N_k(N-1)} = \frac{N_k-1}{N-1}.  \label{eq:tieratio}
\end{equation}
Positions with a greater ratio of internal:total ties than this value prefer to communicate among themselves. Positions with a smaller fraction of internal:total ties than (\ref{eq:tieratio}) prefer to communicate ``outward.'' By also considering whether the position tends to receive ties (proportion received $\sim0$ or $>0$), a social function for the group can be approximated. \citet{burt_positions_1976} names these groups Isolate (prefer internal ties, $\sim0$ received), Primary (prefer internal, $>0$ received), Sycophant (prefer external, $\sim0$ received), and Broker (prefer external, $>0$ received). ``Sycophant'' is a pointlessly negative term in class settings where students are encouraged to seek help from peers, so we might replace it with Advice-Seeker. 

When we calculate the tie ratios and received links for each position in the week 1 and week 10 networks, we find that the ``island'' positions shown on many of the reduced networks are actually a mixture of Primary and Isolate types. (For true isolates---dark blue nodes on the socigrams---the ratio to compare to (\ref{eq:tieratio}) is undefined because there are no ties, internal or otherwise. We include them with the Isolate classification.) The results by class type were: 

\begin{itemize}
	\item Peer Instruction: Both week 1 and week 10 have four Primary positions (preferring within-block interactions, incoming ties) and one Isolate.
	\item SCALE-UP: Week 1 has one Primary (block 1) and four Isolates (blocks 2-5). Week 10 has four Primaries.
	\item ISLE: Week 1 has one Broker (incoming ties but low internal communication), one Primary, and three Isolates, in that order. Week 10 has four Isolates (blocks 1 and 2 have only one incoming link each).
	\item Context-rich problems: Week 1 has one Primary, one Broker, and three Isolates. Week 10 has four Primaries.
\end{itemize}
Figure 9 in the Supplemental Materials \citep{suppMat} shows the link ratios and counts of incoming links for each position. 

Considering tie ratios and incoming links adds nuance to the reduced network plots (Figs.\ \ref{fig:PI-networks}, \ref{fig:SU-networks}, \ref{fig:IS-networks}, and \ref{fig:CR-networks}). In many cases, the number of incoming ties was small enough to be thresholded out of the reduced network display, but is not actually zero. In the SCALE-UP section, an early-semester trend of Isolate positions became a late-semester tendency for Primary positions, reflecting a higher level of overall connectivity even as subgroups gained coherence. On the other hand, for ISLE, the late-semester ``islands'' on the reduced network have very little between-position communication. 
The context-rich problems section, despite its many partially or completely unobserved ties, has a large amount of between-position traffic. Somewhat surprisingly, none of the positions in any network are identified as Advice-Seekers, usually because their amount of internal communication is too high.

\section{Conclusions}\label{sec:conclus}

The goal of this investigation was to compare the network positions available in four active learning classroom types using the method of positional analysis. This technique, which we have not seen used in PER, provides a kind of mid-scale description of group social structure. We structured this investigation around three research questions, discussed below:

\paragraph*{1. What network positions emerge from the four different curriculum types?}
As detailed in section \ref{subsec:positions} and Fig.\ 9 of the Supplemental Materials, most of the node blocks identified by CONCOR preferred links among themselves, but often had at least some level of incoming links, leading to a mixture of Isolate and Primary positions. The image matrices (0/1 representations of the reduced networks) showed a corresponding tendency toward the ``coherent subgroups'' type, where most positions were connected primarily among themselves. 

\paragraph*{2. What differences exist between early- and late-term network positions?}
The week one networks had more deviations from the pattern described above, with the blockmodel plots generally showing more distinct and coherent blocks in week ten. The reciprocity of links also increased from early to late semester, as did the density and average degree (though not always significantly, see Table \ref{tab:nwdesc}). This increase in network connectivity is likely a mixture of effects: social connections forming through the class, but also possibly more students knowing each other's names by the time of the late-semester network survey. 

\paragraph*{3. What major similarities or differences exist in network positions across learning environments?} At the level of analysis provided by CONCOR, there appear to be more similarities than differences in the network position structure. From the data available here, we might say that the ``signature'' of active learning at the positional analysis level is one of coherent subgroups plus a handful of inter-position links. Only a small number of students, if any, are true isolates (degree 0). 

The mixture of Isolate and Primary positions, which can both appear as self-connected ``islands'' on reduced networks, is interesting. The distinction is not trivial---some positional analysis approaches argue that any link at all between positions should be regarded as a positive tie on the image matrix, and a complete absence of links is the most significant structural cue to look for \citep{white_social_1976}. If this stricter standard were to be applied, the reduced networks would be more interlinked, and only a small number of self-linked islands would remain.

Reduced networks are a representation that is uncommon outside of positional analysis, though there are some similar ideas in the flow diagrams of the Infomap algorithm \citep{rosvall_maps_2008} or in node-consolidating analyses of large networks \citep{stanley_compressing_2018}. They lose much of the detail of the full sociogram, but permit comparing position structure between networks of different sizes. We also found that the reduced networks were relatively stable to missing data (details in Supplemental Material \citep{suppMat}). We would not recommend reduced networks as the only description of a network, because the nuance they lose is important to many research questions. But in our data, they show broad structural similarities between classroom networks of different sizes and densities. 

Active learning classrooms often encourage students to work in small groups or pairs, so it is perhaps not surprising that the class-level network structure is most closely classified as coherent subgroups. However, many social interactions that happen in groups of that size lead to different larger-scale structures. Classic blockmodeling studies often find unofficial structures of authority, deference, or a center/periphery structure \citep{white_social_1976}. CONCOR studies of the world trade network have explored its core-periphery structure, reinforcing that the coherent subgroups pattern is not simply an artifact of the algorithm. If a block of students was systematically more popular under the ``meaningful interaction'' prompt, in an unreciprocated way, the reduced networks would show a more star-like structure with single-direction links going to a central hub. In comparison to those findings, the social structure of the classes we surveyed appears to be ``flatter,'' with less hierarchical tendency for the ties between positions. 

Positional analysis provides a different lens for examining networks: more detail than whole-network statistics like average degree or centralization, but  abstracting away some of the node-level detail of centrality scores. If centrality analyses ask ``who has the most power?'', positional analyses ask ``what is the terrain?'' Both characterizations are valuable for understanding the complex system of interactions in a classroom. 

One thing that is not obvious from this analysis is whether there is a preferred or ideal position structure for an active learning environment. Anecdotally, when discussing network analysis with instructors for the first time, a common expectation is that ``everyone will just work with the smart student.'' If it were the case that students informally identified the student(s) most likely to know the correct answers, and worked with those people exclusively, we might see a hierarchy pattern on the image matrix: a few students occupying a position that all other positions directed links toward, creating a single column and row with 1's and 0's in most or all other spaces. This pattern, which did not appear in our data, might indicate an inefficient use of the opportunities in group collaboration. While a few students will often have more prior knowledge and practice with a topic, even initially high-scoring students benefit from collaborating \citep{heller_teaching_1992a}, so a strong hierarchy is unlikely to be a desirable signature for active learning environments. Beyond that, other factors are likely to intervene, such as the physical layout of the classroom and the social structure encouraged by the curriculum. For example, Peer Instruction directs students to talk to a neighbor, and in a large lecture hall with immobile seating, this may tend to lead to ``chains'' in the network rather than large interconnected clusters \citep{commeford_characterizing_2020}. The coherent subgroups signature here, when checked against the sociograms, often summarizes a collection of higher-density groups with a smaller number of ties between them. For this study, we chose project sites with expert implementations of successful active learning curricula, so we argue that the coherent subgroups pattern is one (though not necessarily the only) signature of a well-functioning active learning environment.

\section{Limitations and Future Work}\label{sec:future}

One of the substantial challenges of this study was response rate: even though instructors endorsed the survey data being collected in their classes, the invitation to participate came from an outside researcher, and this may have contributed to the lower response in many sections. The highest-response section in the data was one where the instructor allocated a few minutes of class time for students to do the survey, but we did not ask for this broadly to keep the ``cost'' to instructors low. Ideally, structural analyses of directed networks would use data with a response rate of 70\% or higher, which was unfeasible in our sample. We were able to somewhat check for the severity of this effect by comparing the all-response with complete cases data (see Supplemental Material), but better estimates ultimately must come from more complete sampling of the network. Instructors collecting data in their own classes tend to see higher return rates, so we hope that other researchers may take up this analysis method for their own classrooms. The R code we developed for the CONCOR algorithm is publicly available \citep{suda_CONCOR_2019}, including plotting reduced networks and image matrices. We anticipate that the coherent subgroups pattern we saw in many cases may be common across a wider range of active learning physics classes. 

Because of the difficulty of securing permission to collect identifiable student data for a multi-site study, the CALEP project focused on classroom observation and network survey data. Outcome information such as pass/fail rates, concept inventory scores, or other measures of learning would be valuable additions to the analysis, as would demographic data such as gender or race and ethnicity. Node-level information about student demographics would allow looking for effects such as homophily, the tendency to socially group with others we perceive as ``like'' us \citep{mcpherson_birds_2001,bruun_time_2014}. 

Finally, networks with multi-relation or longitudinal data are especially good candidates for positional analysis. By combining snapshots of the connections between nodes and the pattern of connection between positions, more complicated patterns of interlocking roles can be extracted \citep{white_social_1976,wasserman_social_1994}. PER studies with access to such data \citep{bruun_talking_2013,vargas_correlation_2018} might benefit from this additional layer of analysis. 

By now, there are a number of descriptive results of student networks in introductory physics courses. To move beyond this stage of collecting baselines and into more inferential and predictive questions, network analysis in PER needs careful survey design, reasonably standard data collection protocols, and community discussion about the models and measures most appropriate to capture the interactions of active learning.

\begin{acknowledgments}
	We are grateful to the instructors who allowed us to observe and survey their courses. This work was supported in part by the National Science Foundation under awards DUE-1711017 and DUE-1712341. 
\end{acknowledgments}
	
\bibliography{networks_classrooms}
	
\end{document}


\title{Supplemental Material}
\maketitle

\section{Respondents-only networks}

Missing responses can seriously distort network data \citep{smith_structural_2013}, including block assignments and even blockmodel structure \citep{znidarsic_non-response_2012}. For unknown networks, it can be difficult to estimate the seriousness of the distortion. Strategies for dealing with missing data include ignoring it (using all available information and treating the rest as missing), reducing the data to fully-observed ties, or imputing missing ties using information that is available. \citet{znidarsic_non-response_2012} tested several strategies for handling missing data and found that in most cases, either ignoring missingness or restricting to fully observed ties caused less distortion in the extracted blockmodel. We have opted to show all available data in the main text of this paper, but repeated the analysis keeping only fully observed ties (``complete cases'') for comparison. The results of the complete cases analysis are below.

\subsection{Network descriptive statistics}

Table \ref{tab:nwdesc} shows the calculated descriptive statistics for the complete cases networks, paralleling Table II in the main text. Reciprocity is identical, since it is calculated only between respondents. Density values should not be directly compared, because the all-links and complete cases versions of a network are different sizes. Average degree is higher in the all-response networks than for complete cases, but this difference is generally within the error bars (with the marginal exception of early term context-rich problems). Transitivity is higher in the complete cases networks, though the difference is only significant for early-term SCALE-UP. 

These trends can be predicted by inspecting the all-response network diagrams in the main text. Average degree sums the rows and columns of the adjacency matrix and divides by the number of nodes, so it is simply $2N_e/N$. In most networks, at least a few non-respondents are named by more than one person, so the numerator of average degree will decrease faster than the denominator. The transitivity of a network is based on the ratio of complete triangles to all connected triples of vertices \citep{newman_structure_2003} (the direction of links is ignored for this ratio). Non-respondents will make unconnected triangles unless they are named by two respondents who also talk to each other. Thus, removing non-respondents tends to remove ``dangling'' nodes and increase the fraction of complete triangles. 

\begin{table*}[h] 
	\caption{Network statistics for the respondents-only networks: number of nodes ($N$) and edges ($N_e$), fraction of reciprocated ties, network density, average degree, and transitivity. For the last four values, the standard error of the final digit is given in parentheses.\label{tab:nwdesc}}
	\begin{ruledtabular}
		\begin{tabular}{llrrrrrrr}
			Site & Section & $N$ & $N_e$ & Reciprocity & Density & Avg.\ Degree & Transitivity\\ 
			\hline
			Peer Instruction & Early &  81 & 104 & 0.63 & 0.016(4) & 2.6(3) & 0.2(1) \\ 
			Peer Instruction & Late &  76 & 117 & 0.70 & 0.021(4) & 3.1(3) & 0.3(1)  \\ 
			SCALE-UP & Early &  29 &  30 & 0.60 & 0.04(1) & 2.1(4) & 0.6(2) \\ 
			SCALE-UP & Late &  41 & 107 & 0.65 & 0.07(1) & 5.2(5) & 0.5(1) \\ 
			ISLE & Early &  14 &  12 & 0.50 & 0.07(3) & 1.7(4) & 0.4(3) \\ 
			ISLE & Late &  16 &  30 & 0.93 & 0.13(4) & 3.8(6) & 0.7(2) \\ 
			Context-rich problems & Early &  27 &  24 & 0.42 & 0.03(1) & 1.8(3) & 0.3(2) \\ 
			Context-rich problems & Late &  20 &  45 & 0.62 & 0.12(4) & 4.5(7) & 0.3(1) \\ 
		\end{tabular}
	\end{ruledtabular}
\end{table*}

\newpage

\subsection{Peer Instruction} 

Reducing to complete cases for Peer Instruction means removing 13 non-respondent nodes and 26 partially-observed ties in week 1, and 21 nodes and 42 ties in week 10.

\begin{figure}[h]
	{\centering
	\includegraphics[width=0.9\linewidth]{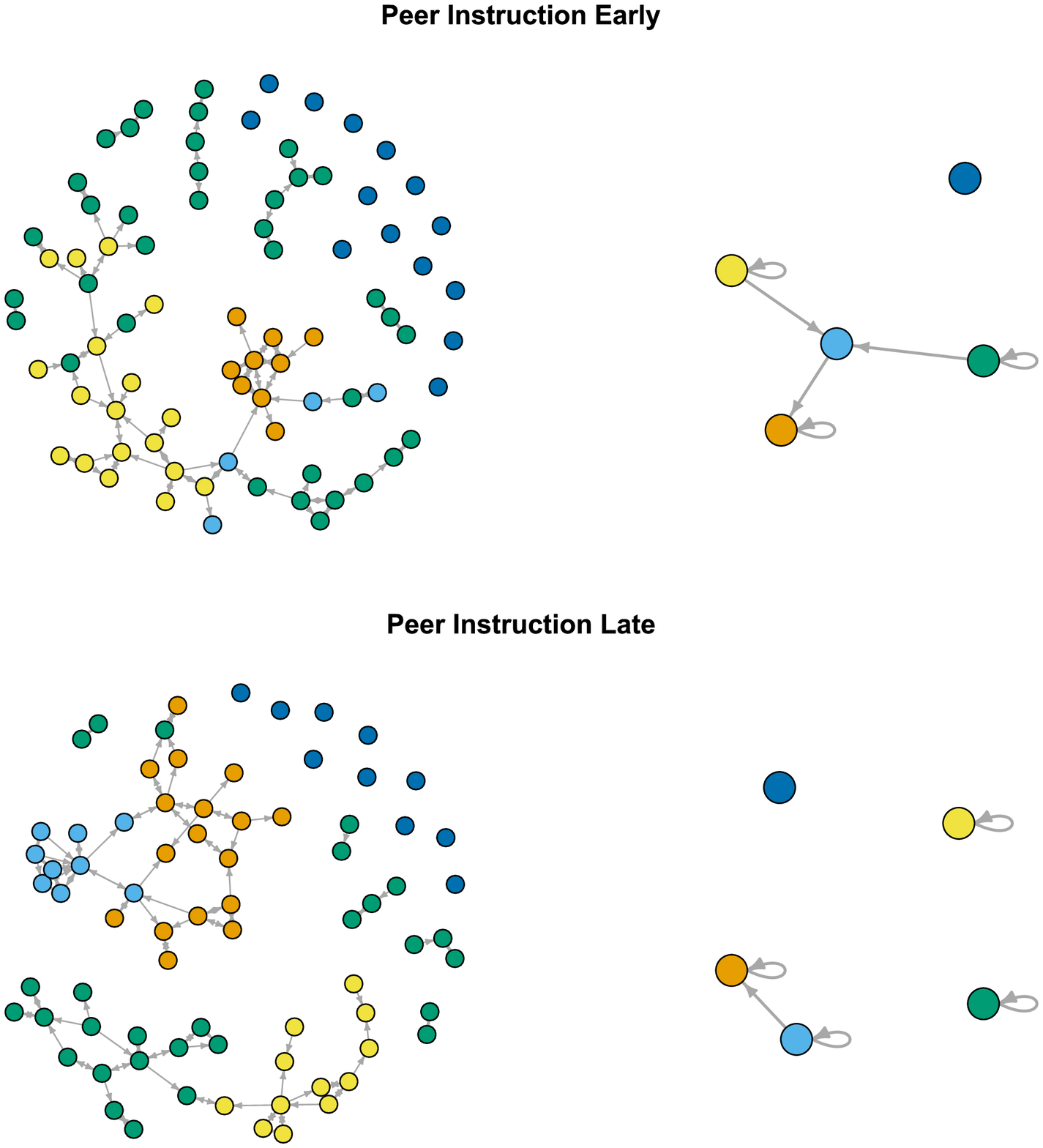} }
	\caption{Week 1 (top) and week 10 (bottom) sociograms and reduced network diagrams for survey respondents in the Peer Instruction section. Nodes are colored by CONCOR partition. Each circle on the reduced network represents a position from the blockmodel and stands for all of the nodes of the same color on the sociogram. Links show connections within and between positions. \label{fig:PI-nws}}
\end{figure}

\begin{figure}[h]
	\includegraphics[width=0.95\linewidth]{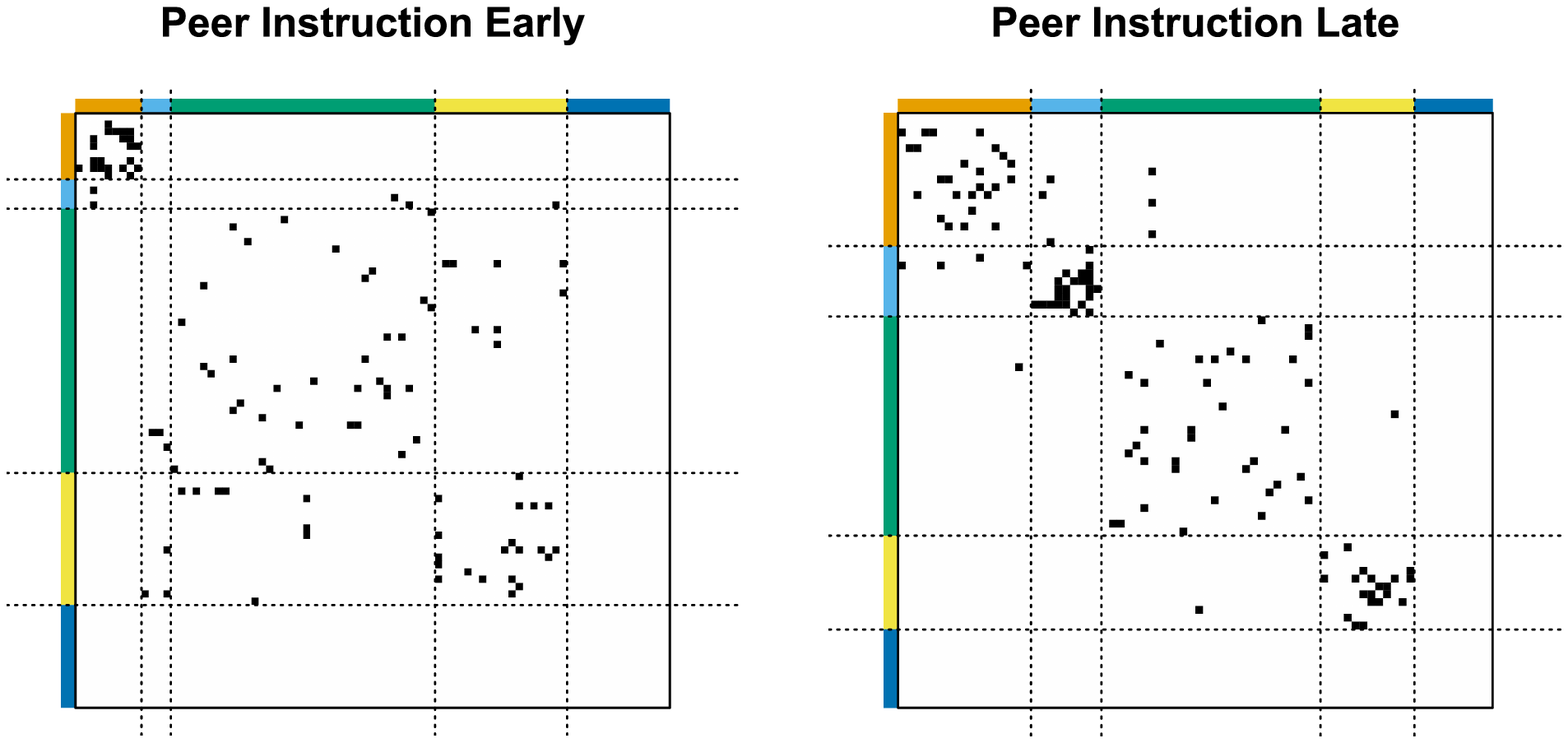}
	\caption{Week 1 (left) and week 10 (right) blockmodels for survey respondents in the Peer Instruction section. Each link from the survey is a black square. The dotted lines mark the CONCOR group partitions. \label{fig:PI-BM}}
\end{figure}

The loss of non-responding nodes changes the early-term reduced network for Peer Instruction substantially. In the all-response network (Fig.\ 3, main text), the giant component consists of two fairly dense blocks with high inner connectivity (orange and light blue) and a somewhat larger and more diffuse block (yellow), while the largest block (green) is primarily nodes on the periphery of the giant component or smaller groups who are not linked to the main mass. In the complete cases network, many nodes from the second position (light blue) are gone and the remaining ones do not directly link to each other, so it is now only a small fraction of the giant component. The first and fourth blockmodel positions (orange and yellow) largely consisted of survey respondents to begin with, so their size and cohesiveness in the giant component are much less affected. Finally, the third blockmodel position (green) still contains smaller connected components, but now has a greater share of the giant component. 

The early-term reduced networks are quite different because of these changes, with the second position now appearing highly connected. This is a side effect of its small size: with so few nodes in that position, even a small number of incoming or outgoing links is likely to rise above the density threshold and appear as a link on the reduced network. 

The late-term block assignments are not as affected by the loss of data. On the sociogram, the giant component was disconnected by the removal of non-respondent nodes, but retains the general pattern of positions (yellow and green are adjoining but with few links between them, light blue and orange have a similar relation but with more blue $\rightarrow$ orange traffic). The week 10 reduced networks are identical between the all-response and complete cases versions. 

\FloatBarrier

\subsection{SCALE-UP}

Reducing to complete cases for SCALE-UP means removing 27 non-respondent nodes and 38 partially-observed ties in week 1, and 25 nodes and 69 ties in week 10.

On the week 1 all-response sociogram, the giant component held members of three positions (orange, light blue, and green), while in the complete-cases network it only includes the first two positions. Several nodes from the third position (green) have been reclassified into one of the first two positions, so that position three only has three members. The fourth position (yellow) is the least affected, having lost three peripheral nodes but otherwise still representing a tightly-connected subgroup not linked to the giant component. These changes are not evident on the reduced network, which is identical between the all-response and complete cases networks.

In week 10, half of the removed nodes come from position four (yellow), so that it loses many of its peripheral nodes. In the all-response network, position four occupies several distinct clusters that are peripheral to or separate from the giant component. In the complete cases network, substantial reassignment of nodes has occurred; position four is now more central in the giant component, and position two (light blue) has taken more of the peripheral nodes. The all-responses reduced network is comprised of coherent subgroups (four self-connected ``island'' positions), but the complete cases network has an isolate position (two nodes who became disconnected when non-respondents were removed) and has enough links from position four to position three (yellow to green) to show on the reduced graph. Its structure still most resembles coherent subgroups, but with some additional complexity because of the holes left in the network by removing non-respondents. 

\begin{figure}[h]
	{\centering
	\includegraphics[width=0.9\linewidth]{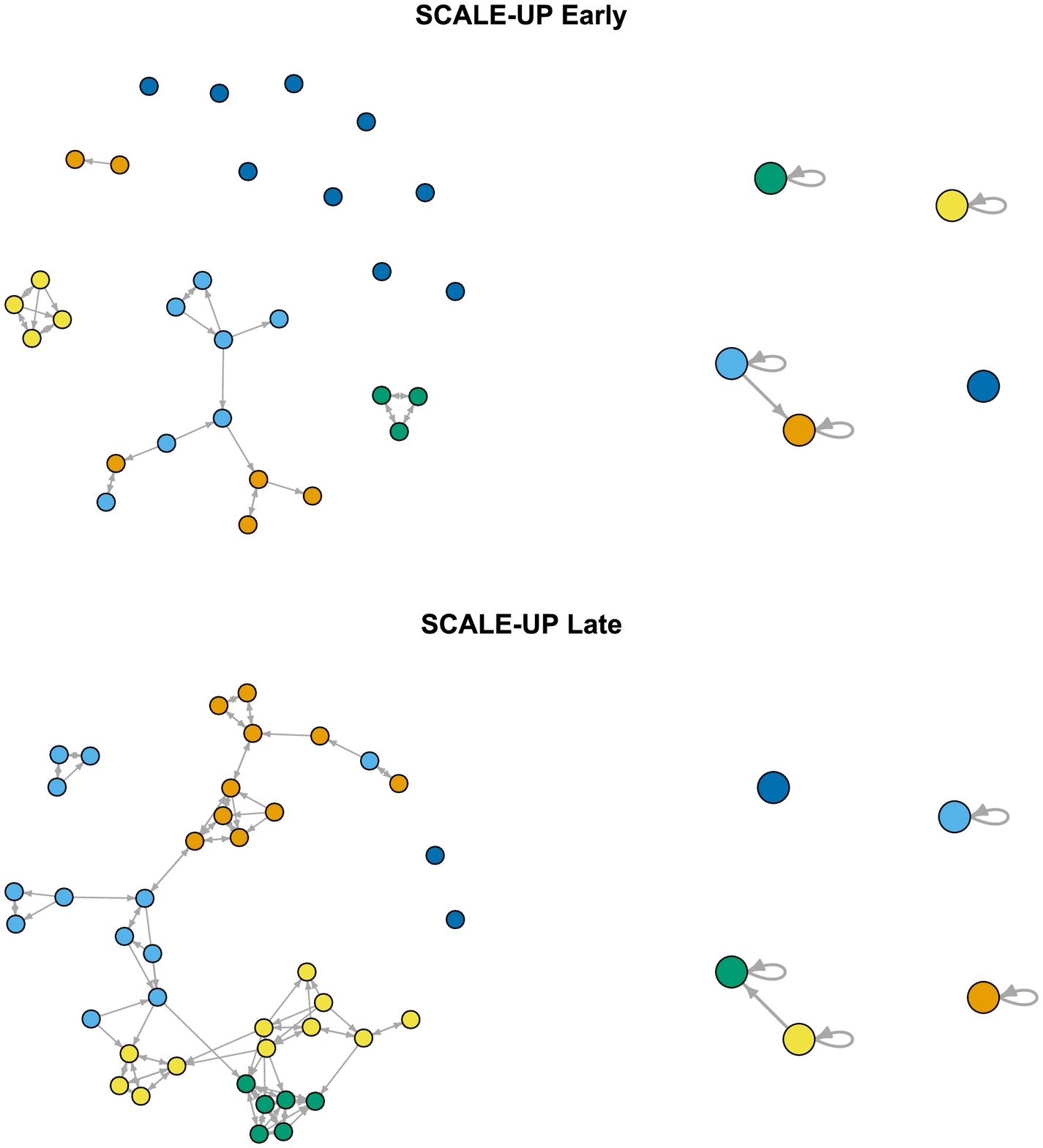} }
	\caption{Week 1 (top) and week 10 (bottom) sociograms and reduced network diagrams for survey respondents in the SCALE-UP section. Nodes are colored by CONCOR partition. \label{fig:SU-socio}}
\end{figure}

\begin{figure}[h]
	\includegraphics[width=0.95\linewidth]{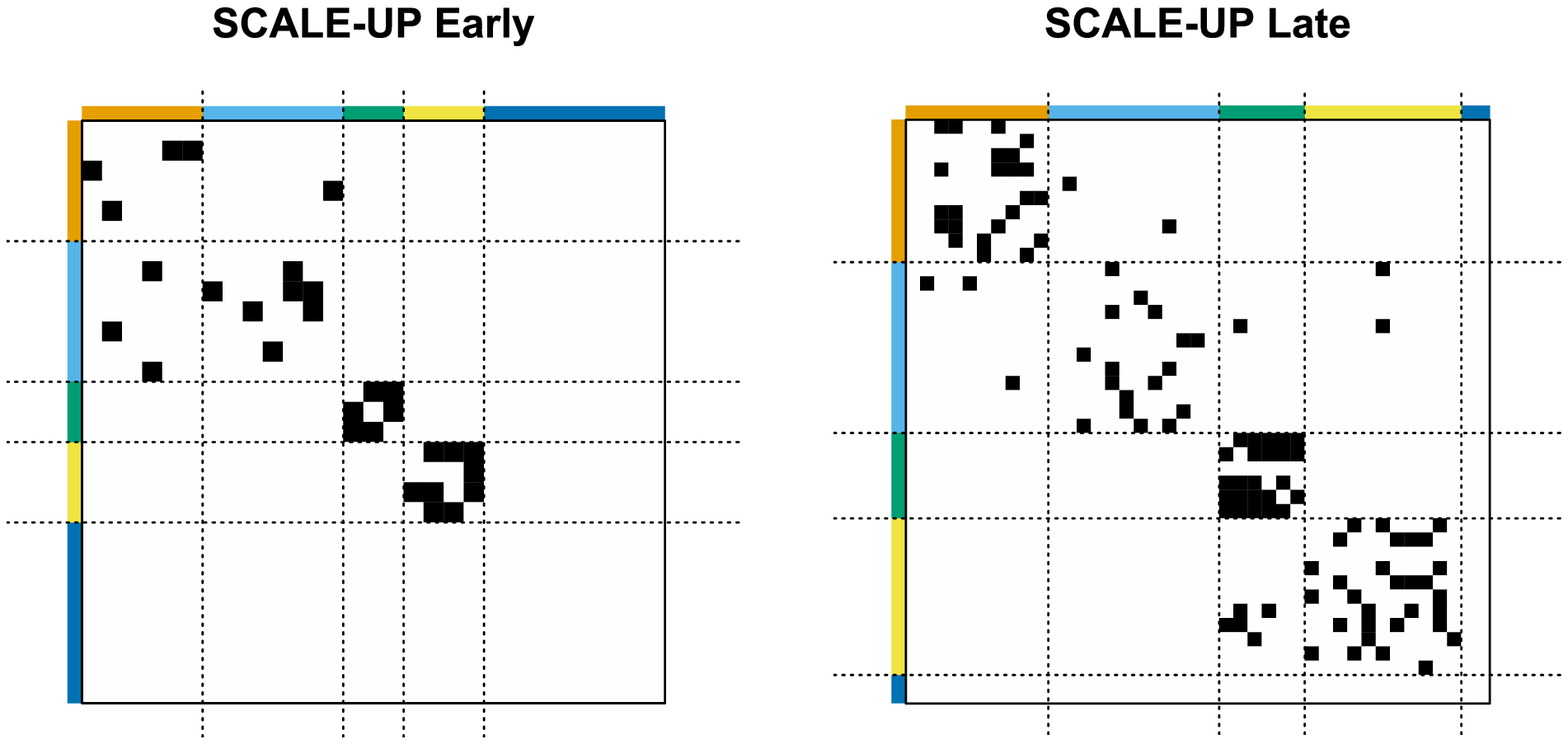}
	\caption{Week 1 (left) and week 10 (right) blockmodels for survey respondents in the SCALE-UP section. Each link from the survey is a black square. The dotted lines mark the CONCOR group partitions. \label{fig:SU-BM}}
\end{figure}

\newpage

\subsection{ISLE}

Reducing to complete cases for ISLE means removing 6 nodes and 7 ties in week 1, and 8 nodes and 17 ties in week 10. 

This was the smallest of the networks, and in some ways it is less affected by the removal of non-respondent nodes. In week 1, on both the all-response and complete cases sociograms, the largest component holds positions one and two (orange and light blue), with smaller disconnected groups for positions three and four, as well as one or two isolates. The reduced networks are identical.

In week 10, removal of non-respondents disconnects the giant component, which in the all-responses network held positions one, two, and four. The largest component in the complete cases network holds positions three and four, with positions one and two now covering free-standing smaller clusters of nodes. However, the reduced networks are identical, with the addition of an isolate position in the complete cases data. 

\begin{figure}[h]
	{\centering
	\includegraphics[width=0.9\linewidth]{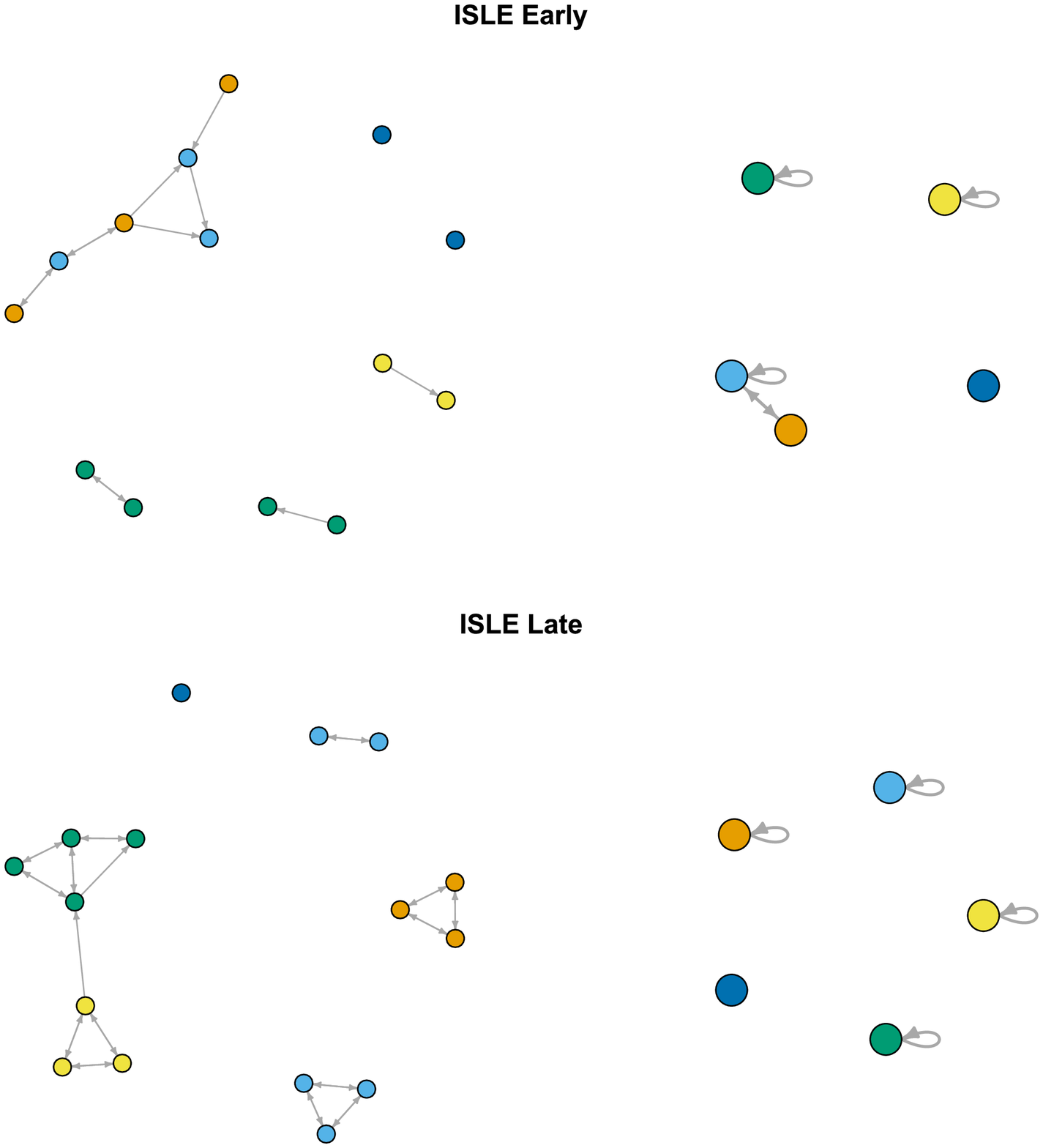} }
	\caption{Week 1 (top) and week 10 (bottom) sociograms and reduced network diagrams survey respondents in the for ISLE section. Nodes are colored by CONCOR partition. \label{fig:IS-socio}}
\end{figure}

\begin{figure}[h]
	\includegraphics[width=0.95\linewidth]{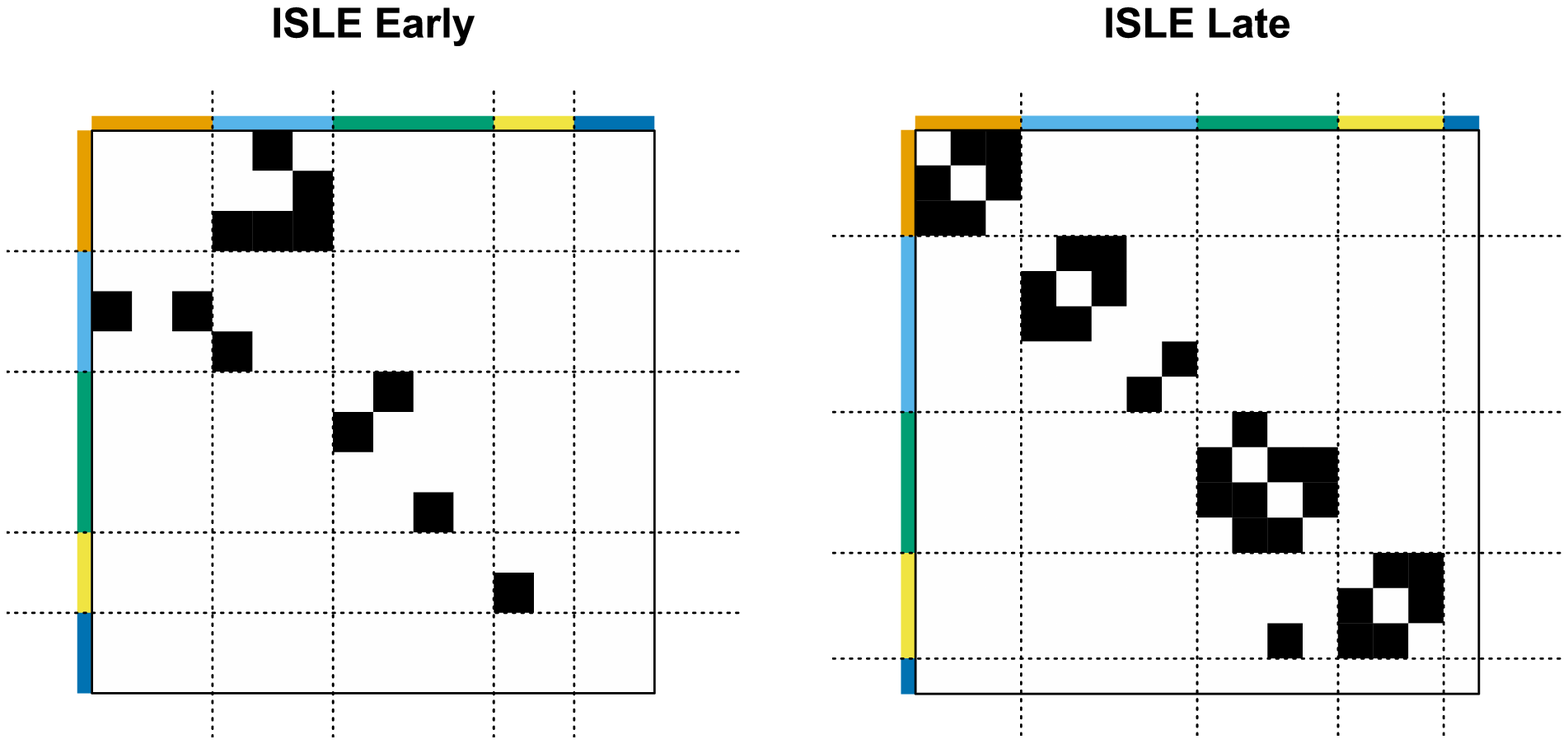}
	\caption{Week 1 (left) and week 10 (right) blockmodels for survey respondents in the ISLE section. Each link from the survey is a black square. The dotted lines mark the CONCOR group partitions. \label{fig:IS-BM}}
\end{figure}

\newpage

\subsection{Context-rich problems}

Reducing to complete cases for context-rich problems means removing 13 nodes and 24 ties in week 1, and 21 nodes and 51 ties in week 10. 

In week 1, restricting to complete cases severs four nodes from the giant component and leaves it with a generally ``stringy'' structure with few triangles. The assignment of nodes has also changed substantially, with position four in complete cases holding several smaller disconnected components, and positions one through three making up the giant component. The link structure of the reduced network is the same, but the position designations (colors) have permuted relative to the all-responses network. 

\begin{figure}[h]
	{\centering
	\includegraphics[width=0.9\linewidth]{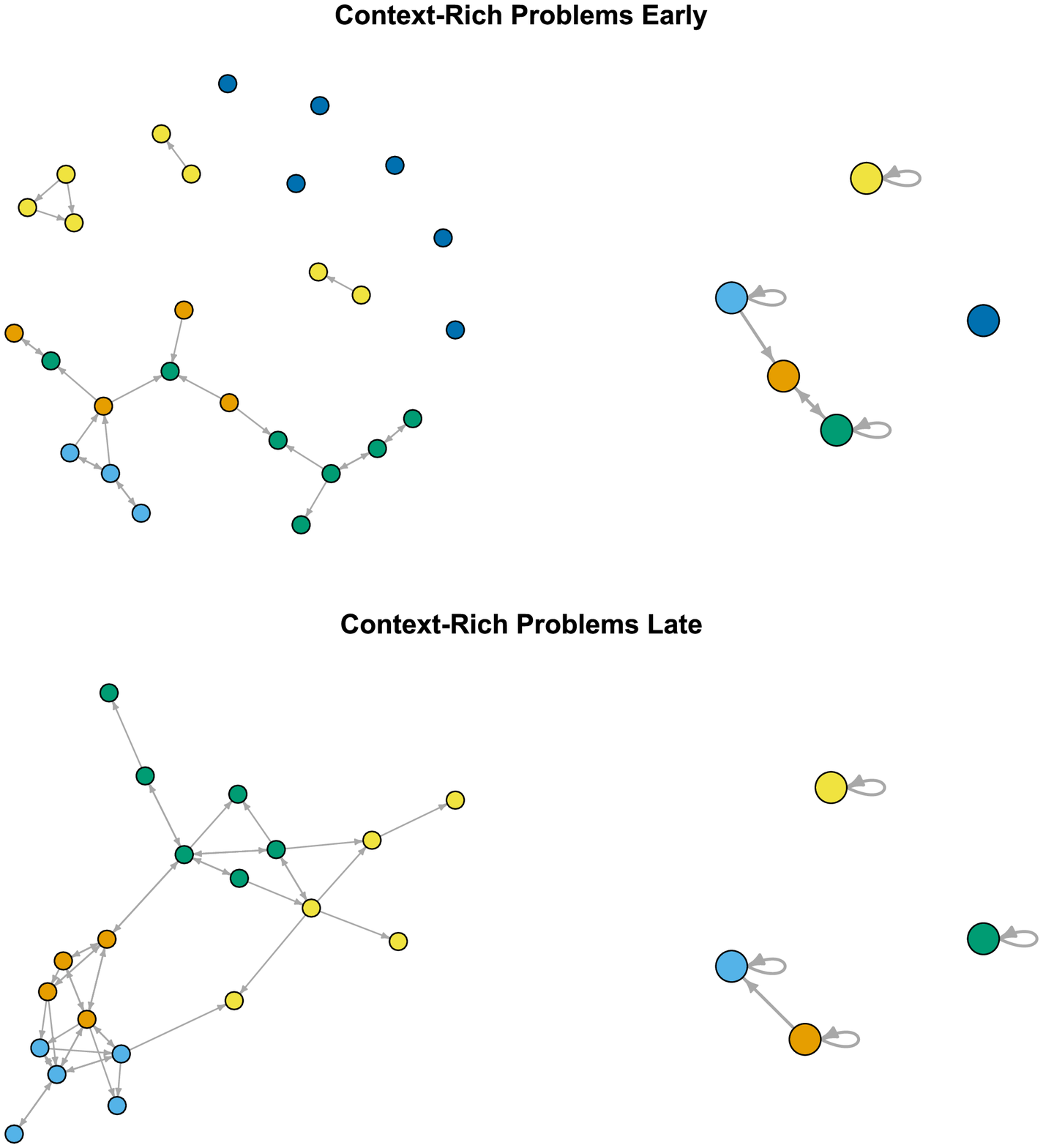} }
	\caption{Week 1 (top) and week 10 (bottom) sociograms and reduced network diagrams for survey respondents in the context-rich problems section. Nodes are colored by CONCOR partition. \label{fig:CR-socio}}
\end{figure}

In week 10, both the all-response and the complete cases networks show all nodes belonging to a single connected component. Removing non-respondents removes a considerable amount of cross-connection in that component, leaving the network in Fig.\ \ref{fig:CR-socio} with four easily distinguished subparts. This segregation is mirrored on the reduced network, which has one fewer between-position link in the complete cases data, and one more within-position link (for position 1, colored orange). 

\begin{figure}[h]
	\includegraphics[width=0.95\linewidth]{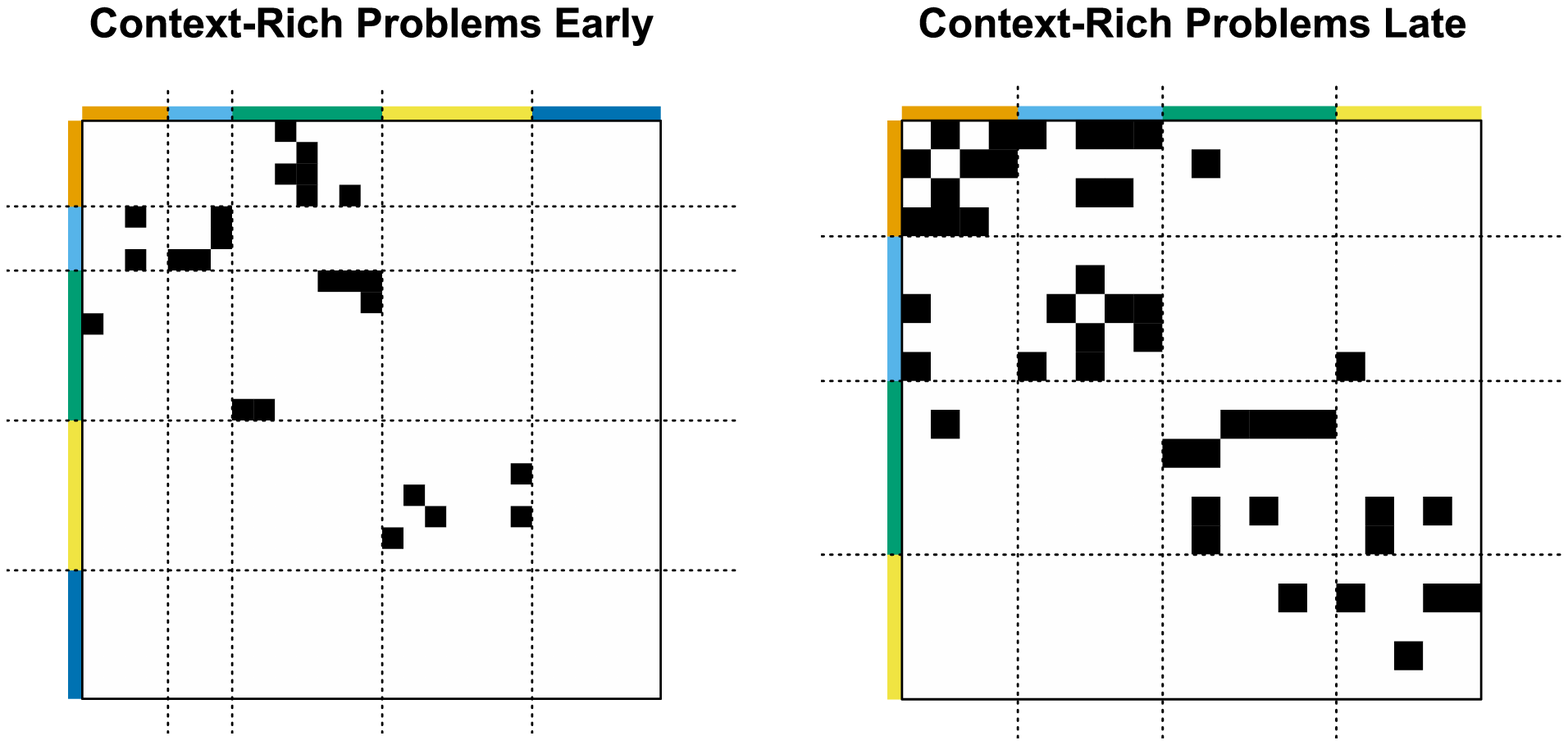}
	\caption{Week 1 (left) and week 10 (right) blockmodels for survey respondents in the context-rich problems section. Each link from the survey is a black square. The dotted lines mark the CONCOR group partitions. \label{fig:CR-BM}}
\end{figure}

\subsection{Differences between all-response and complete cases representations}

Sociograms and reduced networks show different ``zoom'' levels of detail for a network. The full sociogram can be difficult to interpret at a glance, while reduced networks sacrifice detail for simplicity. Comparing the figures above with their counterparts in the paper (Figs. 3, 5, 7, and 9) shows an additional feature of reduced networks. Removing non-respondent nodes, in many cases, substantially fragments the sociogram. Some of these changes are reflected in the reduced networks, but to a lesser degree. The density threshold used for links on the reduced network means that two smaller positions on the reduced network can stay linked, even if their appearance on the sociogram is substantially more tenuous. In that sense, the reduced networks in our sample were somewhat more robust to missing data than the full networks, and thus are a good companion representation to the sociograms. 

\newpage

\section{Link ratios between positions}

\begin{figure*}
	\includegraphics[width=\linewidth]{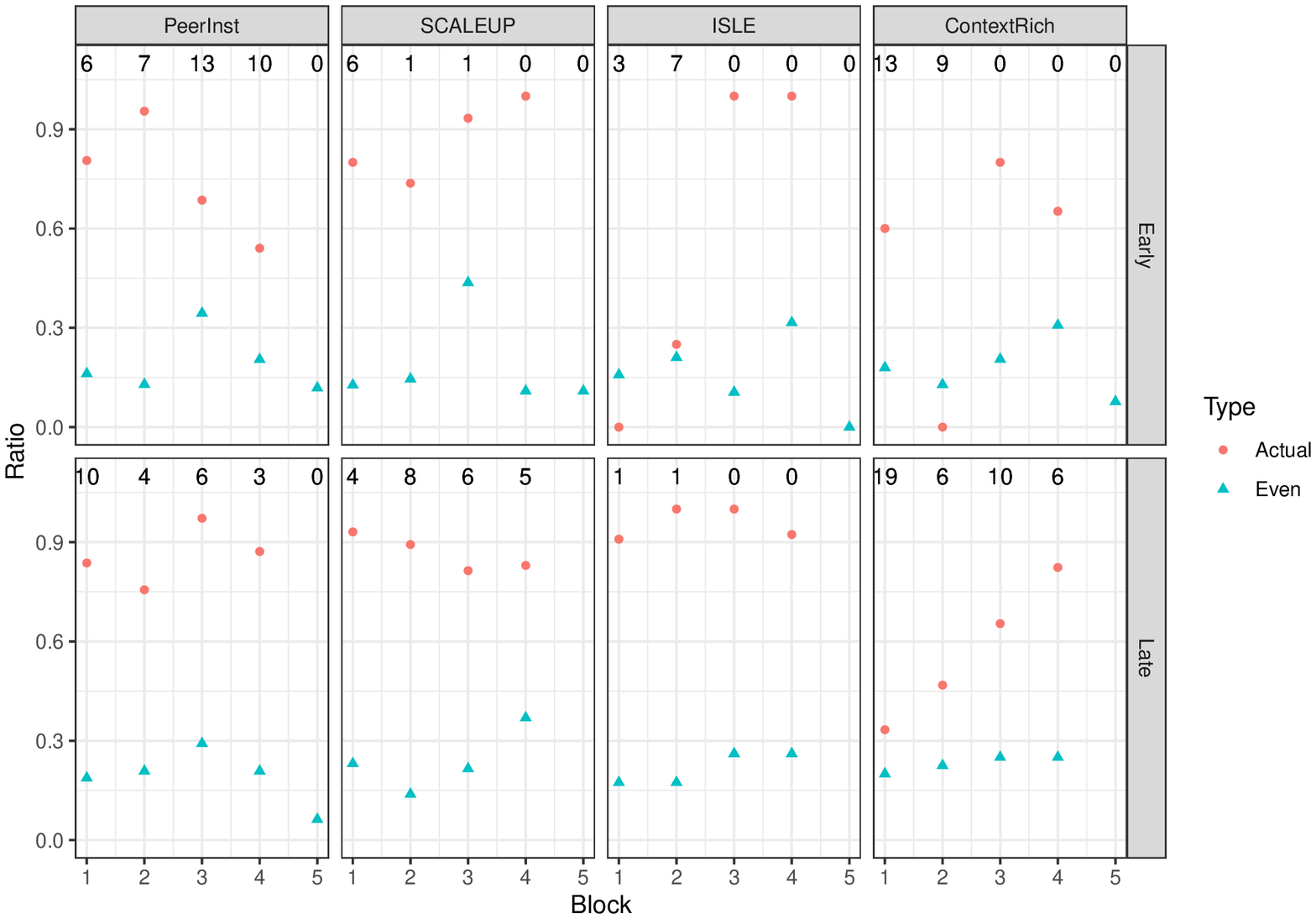}
	\caption{Observed tie ratios (Actual) compared with the values that would show no preference for within-group or external communication (Even). The top row shows week 1 results, and the bottom row shows week 10 results. When present, position 5 indicates the isolates in the network (survey respondents who chose and received no links). Integers at the top of each section show the number of incoming links for each block. \label{fig:tieratio}}
\end{figure*}

Following the discussion in Section IV.B of the paper, we can classify a network position by its preference for internal vs. external communication and the tendency for other positions to link to it. The ``break-even'' tie ratio describing communication preference is 

\begin{equation}
\frac{N_k(N_k-1)}{N_k(N-1)} = \frac{N_k-1}{N-1},  \label{eq:tieratio}
\end{equation}
where $N$ is the number of nodes in the network, and $N_k$ is the number in position $k$ \citep[chapter 10]{wasserman_social_1994}. 
Figure \ref{fig:tieratio} shows the observed tie ratios and the threshold value of (\ref{eq:tieratio}) for each block). In most cases, there is a strong preference for within-block communication---the observed ratio is substantially higher than the ``break-even'' value. At the top of each section display is the number of incoming links received by that block. 

For discussion purposes, positions with 0 or 1 incoming link are classed as not receiving outside communication. This choice means that in smaller sections, such as ISLE, classifying positions as Isolates is more sensitive to small effects from missing data. If a stricter threshold of 0 incoming links is used for the ISLE section, its week 1 descriptions remain unchanged, but week 10 would have two Primary and two Isolate positions. 

\bibliography{networks_classrooms}